\documentclass[longauth]{aa}
\usepackage{float}
\usepackage{graphicx}
\usepackage{txfonts}
\usepackage{hyperref}
\usepackage{url}
\usepackage{natbib}
\usepackage{multirow}
\usepackage{subfig}
\usepackage{supertabular,booktabs}
\usepackage{longtable}
\usepackage{ltxtable} 
\usepackage{tabularx} 
\usepackage{ltablex}
\begin{document}

   \title{Many new variable stars discovered in the core of the globular cluster
NGC~6715 (M54) with EMCCD observations\thanks{Based on data collected by MiNDSTEp with the Danish 1.54 m telescope at the ESO La Silla observatory}}

   \titlerunning{Variable stars in the core of NGC~6715}

   \author{R.~{Figuera~Jaimes} \inst{1,2} \thanks{email: robertofiguera@gmail.com}
      \and D.~M.~Bramich \inst{3} \thanks{email: dan.bramich@hotmail.co.uk}
      \and N.~Kains \inst{6}
      \and J.~Skottfelt \inst{4,5}
      \and U.~G.~J\o{}rgensen \inst{5}
      \and K.~Horne \inst{1}
	  \and M.~Dominik \inst{1} \thanks{Royal Society University Research
Fellow}
	  \and K.~A.~Alsubai \inst{3}
	  \and V.~Bozza \inst{7,8}
          \and M.~J.~Burgdorf \inst{26}
	  \and S.~{Calchi~Novati} \inst{9,7,10} \thanks{Sagan visiting fellow}
	  \and S.~Ciceri \inst{25}
	  \and G.~D'Ago \inst{10}
          \and D.~F.~Evans \inst{21}
	  \and P.~Galianni \inst{1}
	  \and S.-H.~Gu \inst{12,13}
 	  \and K.~B.~W~Harps\o{}e \inst{5}
	  \and T.~Haugb\o{}lle \inst{5}
	  \and T.~C.~Hinse \inst{14}
	  \and M.~Hundertmark \inst{5,1}
	  \and D.~Juncher \inst{5}
          \and E.~Kerins \inst{29}
	  \and H.~Korhonen \inst{24,15,5}
          \and M.~Kuffmeier \inst{5}
	  \and L.~Mancini \inst{11}
          \and N.~Peixinho \inst{27,28}
	  \and A.~Popovas \inst{5}
	  \and M.~Rabus \inst{16,11}
	  \and S.~Rahvar \inst{17}
	  \and G.~Scarpetta \inst{10,7,8}
	  \and R.~W.~Schmidt \inst{18}
	  \and C.~Snodgrass \inst{19,20}
	  \and J.~Southworth \inst{21}
	  \and D.~Starkey \inst{1}
	  \and R.~A.~Street \inst{22}
	  \and J.~Surdej \inst{23}
          \and R.~Tronsgaard \inst{30}
          \and E.~{Unda-Sanzana} \inst{27}
          \and C.~{von~Essen} \inst{30}
	  \and X.-B.~Wang \inst{12,13}
	  \and O.~Wertz \inst{23}
\\
	  (The MiNDSTEp Consortium)
}

   \institute{SUPA, School of Physics and Astronomy, University of St. Andrews, North Haugh, St Andrews, KY16 9SS, United Kingdom.
    \and European Southern Observatory, Karl-Schwarzschild-Stra\ss{}e 2, 85748 Garching bei M\"{u}nchen, Germany 
    \and Qatar Environment and Energy Research Institute (QEERI), HBKU, Qatar Foundation, Doha, Qatar 
    \and Centre for Electronic Imaging, Dept. of Physical Sciences, The Open University, Milton Keynes MK7 6AA, UK 
    \and Niels Bohr Institute and Centre for Star and Planet Formation, University of Copenhagen, {\O}ster Voldgade 5, 1350 Copenhagen K, Denmark 
    \and Space Telescope Science Institute, 3700 San Martin Drive, Baltimore, MD 21218, United States of America 
    \and Dipartimento di Fisica ''E. R. Caianiello'', Universit\`a di Salerno, Via Giovanni Paolo II 132, 84084-Fisciano (SA), Italy 
    \and Istituto Nazionale di Fisica Nucleare, Sezione di Napoli, Napoli, Italy 
    \and NASA Exoplanet Science Institute, MS 100-22, California Institute of Technology, Pasadena CA 91125 
    \and Istituto Internazionale per gli Alti Studi Scientifici (IIASS), 84019 Vietri Sul Mare (SA), Italy 
	\and Max Planck Institute for Astronomy, K\"onigstuhl 17, 69117 Heidelberg, Germany 
	\and Yunnan Observatories, Chinese Academy of Sciences, Kunming 650011, China  
	\and Key Laboratory for the Structure and Evolution of Celestial Objects, Chinese Academy of Sciences, Kunming 650011, China 
	\and Korea Astronomy and Space Science Institute, Daejeon 305-348, Republic of Korea 
	\and Finnish Centre for Astronomy with ESO (FINCA), University of Turku, V{\"a}is{\"a}l{\"a}ntie 20, FI-21500 Piikki{\"o}, Finland 
	\and Instituto de Astrof\'isica, Facultad de F\'isica, Pontificia Universidad Cat\'olica de Chile, Av. Vicu\~na Mackenna 4860, 7820436 Macul, Santiago, Chile 
	\and Department of Physics, Sharif University of Technology, P. O. Box 11155-9161 Tehran, Iran 
	\and Astronomisches Rechen-Institut, Zentrum f\"ur Astronomie der Universit\"at Heidelberg, M\"onchhofstr. 12-14, 69120 Heidelberg, Germany 
    \and Planetary and Space Sciences, Department of Physical Sciences, The Open University, Milton Keynes, MK7 6AA, UK 
    \and Max-Planck-Institute for Solar System Research, Justus-von-Liebig-Weg 3, 37077 G\"ottingen, Germany 
	\and Astrophysics Group, Keele University, Staffordshire, ST5 5BG, UK 
	\and Las Cumbres Observatory Global Telescope Network, 6740 Cortona Drive, Suite 102, Goleta, CA 93117, USA 
	\and Institut d'Astrophysique et de G\'eophysique, Universit\'e de Li\`ege, All\'ee du 6 Ao\^ut 9c, 4000 Li\`ege, Belgium 
        \and Dark Cosmology Centre, Niels Bohr Institute, University of Copenhagen, Juliane Maries vej 30, DK-2100 Copenhagen {\O}, Denmark 
        \and Department of Astronomy, Stockholm University, AlbaNova University Center, SE-106 91 Stockholm, Sweden 
        \and Meteorologisches Institut, Universit{\"a}t Hamburg, Bundesstra\ss{}e 55, 20146 Hamburg, Germany 
        \and Unidad de Astronom\'ia, Fac. de Ciencias B\'asicas, Universidad de Antofagasta, Avda. U. de Antofagasta 02800, Antofagasta, Chile 
        \and 20 CITEUC - Centre for Earth and Space Science Research of the University of Coimbra, Observat\'orio Astron\'omico da Universidade de Coimbra, 3040-004 Coimbra, Portugal 
        \and Jodrell Bank Centre for Astrophysics, School of Physics and Astronomy, University of Manchester, Oxford Road, Manchester M139PL, UK 
        \and Stellar Astrophysics Centre, Department of Physics and Astronomy, Aarhus University, Ny Munkegade 120, 8000 Aarhus C, Denmark 
}
  \abstract
  {We show the benefits of using Electron-Multiplying CCDs and the shift-and-add 
technique as a tool to minimise the effects of the atmospheric turbulence such as blending 
between stars in crowded fields and to avoid saturated stars in the fields observed. We intend to
complete, or improve, the census of the variable star population in globular cluster 
NGC~6715.}
   {Our aim is to obtain high-precision time-series photometry of the very crowded 
central region of this stellar system via the collection of better angular resolution images than has been 
previously achieved with conventional CCDs on ground-based telescopes.}   
   {Observations were carried out using the Danish 1.54-m Telescope at the ESO La Silla observatory in Chile. The telescope is equipped with an Electron-Multiplying CCD that allowed to obtain short-exposure-time images (ten images per second) that were stacked using the shift-and-add technique to produce the normal-exposure-time images (minutes). The high precision photometry was performed via difference image analysis employing the DanDIA pipeline. We attempted automatic detection of variable stars in the field.}
   {We statistically analysed the light curves of 1405 stars in the crowded central region of NGC~6715 to automatically identify the variable stars present in this cluster. We found light curves for 17 previously known variable stars near the edges of our reference image (16 RR Lyrae and 1 semi-regular) and we discovered 67 new variables (30 RR Lyrae, 21 long-period irregular, 3 semi-regular, 1 W Virginis, 1 eclipsing binary, and 11 unclassified). Photometric measurements for these stars are available in electronic form through the Strasbourg Astronomical Data Centre.}
   {}
   \maketitle
%
\section{Introduction}

Galactic globular clusters are interesting stellar systems in Astronomy as they are 
fossils of the early Galaxy formation and evolution. This makes them excellent 
laboratories in a wide range of topics from Stellar Evolution to Cosmology, from 
observations to theory.

NGC~6715 (M54) was discovered on July 24, 1778 by Charles 
Messier\footnote{http://messier.seds.org/m/m054.html}. The cluster is in the Sagittarius 
Dwarf Spheroidal galaxy at a distance of 26.5 kpc from our Sun and 18.9 kpc from our 
Galactic centre. It has a metallicity [Fe/H]=$-$1.49 dex and a distance modulus 
(m$-$M)$_V$=17.58 mag. The magnitude of its horizontal branch is V$_{\mathrm{HB}}$=18.16 
mag \citep[2010 version][]{harriscatalog}. The stellar population and morphology of NGC~6715
have attracted the attention of several studies. Although NGC~6715’s position is aligned 
with the central region of the Sagittarius Dwarf Spheroidal galaxy, recent studies have 
shown that
this cluster is not the nucleus of the mentioned galaxy \citep{layden00+01, majewski03+03, monaco05+03, 
bellazzini08+08}. \cite{carretta10+09} found that the metallicity spread in this 
cluster is intermediate between smaller-normal Galactic globular clusters and metallicity 
values associated with dwarf galaxies. It has a very peculiar colour-magnitude diagram 
with 
multiple main sequences, turnoff points and an extended blue horizontal branch (HB) 
\citep{siegel07+13, milone15}. \cite{rosenberg04+02} found that this cluster hosts 
a 
blue hook stellar population in its blue HB. It is thought that NGC~6715 might host an 
intermediate mass black hole (IMBH) in its centre \citep{ibata09+10, wrobel11+02} as well.

\cite{oosterhoff39} found that globular clusters can be classified into two 
groups based on the mean periods and the number ratios of their RR0 and RR1 stars. These 
are the Oosterhoff I (OoI) and Oosterhoff II (OoII) groups, and this is known as the 
Oosterhoff dichotomy. It was also found that the metallicity of the clusters plays an 
important role in this classification \citep{kinman59} and that it could be related with 
the formation history of the Galactic halo \citep[see e. g. discussions 
in][]{lee99+01, catelan04, catelan09, smith11+03, sollima14+03}.

The first Oosterhoff classification for NGC 6715 was done by \cite{layden00+01}  
by 
comparing the amplitudes and periods of 67 RR Lyrae in this cluster with the 
period-amplitude relation obtained for M3 (OoI) and M9 (OoII). NGC 6715 follows the 
relation defined by M3 and this implies that it is an OoI cluster. They also found that 
the mean periods of the RR Lyrae also agree with the OoI classification. Similarly, 
\cite{sollima10+03} in their study of variable stars in NGC 6715 discovered a new set of 
80 RR Lyrae, and they used 95 RR0 and 33 RR1 (after excluding non-members or problematic 
stars) to find that this cluster shares some properties of both OoI and OoII clusters. 
For instance, the average period of its RR0 and RR1 stars was found to be intermediate 
between the values for the OoI and OoII classifications. We note that \cite{catelan09} 
also classified this cluster as an intermediate Oosterhoff type based on its position in 
the $<P_{\mathrm{RR0}}>$-metallicity diagram.

Several time-series photometric studies focused on the variable star population 
\citep{rosino52, rosino59+01, layden00+01, sollima10+03, li13+01}, but due to the high 
concentration of stars in its core, many variables have previously been missed due to 
blending. NGC~6715 is very massive and it is among the densest known 
globular clusters \citep{pryor93+01}. Due to this, NGC~6715 also took our 
attention as an excellent candidate for our study of globular clusters using 
Electron-Multiplying CCDs (EMCCDs) and the shift-and-add technique 
\citep{skottfelt13, Skottfelt15+05, figuera15}.

Section \ref{sec:data_red} summarises the observations, the reduction, and photometric 
techniques employed. Section \ref{sec:calib} explains the calibration applied to the 
instrumental magnitudes. Section \ref{sec:var_search} contains the technique used in the 
detection and extraction of variable stars. Sections \ref{sec:var_class} and \ref{sec:cmd} 
show the methodology used to classify variables and the color-magnitude diagrams employed, 
respectively. Section~\ref{sec:cluster_results} presents our results and
Section \ref{sec:oo_type} discusses the Oosterhoff classification 
of this cluster. Our conclusions are presented in \ref{sec:conclusion}. 

\section{Data, reduction, and photometry}\label{sec:data_red}

EMCCDs, also known as low light level charge-coupled devices (L3CCD) 
\citep[see e. g.][]{smith08+06, jerram01+09}, are conventional CCDs that have 
an extended readout register where the signal is amplified by impact ionisation before
they are read out. Hence the readout noise is negligible when compared to the signal
and very high frame-rates become feasible (10-100 frames/s). This provides an opportunity
to compensate for the blurring effect of the turbulence in the atmosphere. By shifting
and adding the individual frames appropriately, it is possible to construct much
higher resolution images than is possible using conventional CCD imaging from the ground.
Furthermore, the dynamic range of the stacked images is greatly increased and the
saturation of bright stars is therefore not an issue except for the very brightest stars in
the sky. However, the main drawback with this technique is that the process whereby the
signal is amplified also increases the photon noise component in the images by a factor of
$\sqrt{2}$ when compared to conventinal CCD images. One should also be aware that EMCCD
imaging data need to be calibrated in a different way to conventional CCD data. Several
studies where EMCCDs have been used can be found in the literature e.g. high-resolution imaging
of exoplanet host stars in the search for unseen companions
\citep{southworth15+mindstep, ciceri16+mindstep, bozza16, strret16, evans16+mindstep} 
photometric and astrometric measurements of a pair of very close brown dwarfs 
\citep{mancini15+mindstep}, and time-series photometry of crowded globular cluster cores
aiming to complete the census of variable stars \citep{Skottfelt15+05, figuera15}.

The data presented in this paper are the result of EMCCD observations performed over three
consecutive years: 2013, 2014 and 2015 in April to September each year. The 1.54 m Danish
telescope at the ESO Observatory in La Silla, Chile was used with an Andor Technology
iXon+897 EMCCD camera, which has a 512 $\times$ 512 array of 16 $\mu$m pixels, giving a
pixel scale of $\sim0^{\prime\prime}$.09 per pixel and a total field of view of
$\sim45\times45$ arcsec$^{2}$.

For this research the EMCCD camera was set to work at a frame-rate of 10 Hz (ten frames 
per second) and an EM gain of $300e^-$/photon. The camera was placed behind a dichroic 
mirror which works as a long-pass filter. Taking the mirror and the sensitivity of the 
camera into consideration, it is possible to cover the wavelength range from 650 nm to 
1050 nm. This is roughly a combination of SDSS $\mathrm{i}^{\prime}+\mathrm{z}^{\prime}$ 
filters \citep{bessell05}. More details about the instrument can be found in 
\cite{Skottfelt15+15}. The total exposure time employed for a single observation 
was 10 minutes, which means that each observation is the result of 
shifting-and-adding 6000 exposures. The resulting PSF FWHM in the reference image
employed in the photometric reductions (see below) was 0.44$^{\prime\prime}$.

In Figure \ref{fig:histograms_nights}, histograms of the number of observations per night 
during each year are shown. Data in the left-hand panel correspond to 2013, data in the 
middle panel to 2014 and data in the right-hand panel to 2015. We aimed to always take two 
observations per night, although it was not always possible due to weather conditions or 
time slots needed for other projects as is the case of the monitoring of microlensing 
events carried out by the MiNDSTEp consortium.

\begin{figure}[htp!]
\centering
\includegraphics[scale=1]{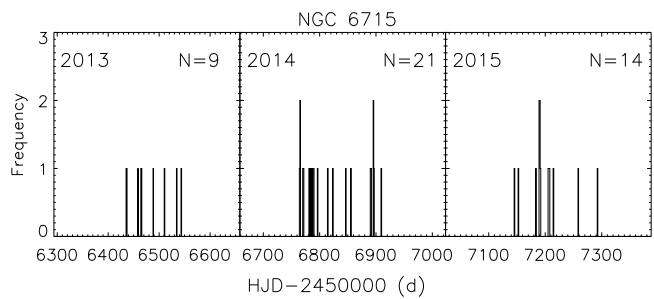}
\caption{Histograms with the number of observations per night for globular
cluster NGC~6715. Left: 2013. Middle 2014. Right: 2015.}
\label{fig:histograms_nights}
\end{figure}

Bias, flat-field and tip-tilt corrections, were performed using the procedures and 
algorithms described in \cite{harpsoe12+03}. In particular, the tip-tilt correction allows 
high-resolution stacked images to be created as described in detail in \cite{Skottfelt15+05, 
figuera15}. Briefly, the method uses the Fourier cross correlation theorem where, for each
bias- and flat-corrected exposure $I_k(i,j)$ within an observation, we calculate the
cross correlation image between $I_k(i,j)$ and an average of 100 randomly chosen exposures.
The peak in the cross correlation image gives the appropriate shift to correct the tip-tilt
error. We then stack the shifted exposures according to batches grouped by image quality
(a measure of image sharpness), to create ten-layer image cubes. The best quality layers
from each cube (i.e. the sharpest layers) are extracted from all of the available observations
to create a high-resolution reference image for subsequent difference image analysis (DIA).
Finally, each ten-layer cube is stacked to create a single image corresponding to each 
observation.

To extract the photometry in each of the stacked images we used the 
DanDIA\footnote{DanDIA is built from the DanIDL library of IDL routines available at 
\url{http://www.danidl.co.uk}} pipeline \citep{bramich08, bramich13+10}, which is based on 
difference image analysis \citep[DIA;][]{alard98+01, alard00}.
The pipeline works by aligning all images to the reference image, solving for a set of convolution
kernels modelled as discrete pixel arrays, and subtracting the convolved reference image in each
case to create a set of difference images. Stars are detected on the reference image and their
reference fluxes $f_{\mathrm{ref}}$ (ADU/s) are measured using PSF photometry. Difference fluxes
$f_{\mathrm{diff}}(t)$ (ADU/s) for each star detected in the reference image are measured in each
of the difference images by optimally scaling the PSF model for the star to the difference image.
The light curve for each star in instrumental magnitudes $m_{\mathrm{ins}}$ are built as 
shown in Eqn.~(\ref{eq:inst_mag})

\begin{equation}\label{eq:inst_mag}
m_{\mathrm{ins}}(t)=17.5-2.5\log(f_{\mathrm{tot}}(t)),
\end{equation}
where $f_{\mathrm{tot}}(t)$ is the total flux in ADU/s defined as 
\begin{equation}\label{eq:total_flux}
f_{\mathrm{tot}}(t)=f_{\mathrm{ref}}+\frac{f_{\mathrm{diff}}(t)}{p(t)}.
\end{equation}

The quantity $p(t)$ is the photometric 
scale factor used to scale the reference frame to each image as part of the kernel model 
explained in \cite{bramich08}.

An electronic table with photometric measurements and fluxes for all the variable stars 
presented in this work is available through the CDS\footnote{http://cds.u-strasbg.fr/} 
database with the format illustrated in Table \ref{tab:electronic_data}.

\begin{table*}[htp!]
\caption{Time-series I-band photometry for all known and new variables in the field of view covered in globular cluster NGC~6715. The standard $M_{\mathrm{std}}$ and instrumental $m_{\mathrm{ins}}$ magnitudes are listed in Columns 4 and 5, respectively, corresponding to the variable star, filter, and epoch of mid-exposure listed in Columns 1-3, respectively. The uncertainty on $m_{\mathrm{ins}}$ is listed in Column 6, which also corresponds to the uncertainty on $M_{\mathrm{std}}$. For completeness, we also list the quantities $f_{\mathrm{ref}}$, $f_{\mathrm{diff}}$ and $p$ from Eqn. \ref{eq:total_flux} in Columns 7, 9 and 11, along with the uncertainties $\sigma_{\mathrm{ref}}$ and $\sigma_{\mathrm{diff}}$ in Columns 8 and 10. This is an extract from the full table, which is available with the electronic version of the article at the CDS.}
\label{tab:electronic_data}
\centering
\tabcolsep=0.195cm
\begin{tabular}{ccccccccccc}
\hline\hline
var & Filter & HJD & $M_{\mathrm{std}}$ & $m_{\mathrm{ins}}$ & $\sigma_m$ &
$f_{\mathrm{ref}}$ & $\sigma_{\mathrm{ref}}$ & $f_{\mathrm{diff}}$ &
$\sigma_{\mathrm{diff}}$ & $p$ \\
id  &        & (d) &       (mag)        &       (mag)        &   (mag)    &
(ADUs$^{-1}$)    &     (ADUs$^{-1}$)     &   (ADUs$^{-1}$)   & (ADUs$^{-1}$) &  
  \\
(1)  &   (2)     & (3) &       (4)        &       (5)        &   (6)    &
(7)    &     (8)     &   (9)   & (10) & (11) \\
\hline
V112 & I & 2456435.88824 & 13.706 & 4.679 & 0.001 & 141855.480 & 1550.259 & -44401.153 & 540.539 & 5.9879 \\
V112 & I & 2456436.89695 & 13.689 & 4.661 & 0.001 & 141855.480 & 1550.259 & -30447.356 & 609.851 & 5.8368 \\
\vdots & \vdots & \vdots & \vdots &\vdots &\vdots &\vdots & \vdots & \vdots & \vdots & \vdots\\
V160 & I & 2456435.88824 & 17.882 & 8.855 & 0.012 & 4069.406 & 1553.328 & -7171.335 & 186.330 & 5.9879 \\
V160 & I & 2456436.89695 & 17.646 & 8.618 & 0.015 & 4069.406 & 1553.328 & -2916.178 & 292.135 & 5.8368 \\
\vdots & \vdots & \vdots & \vdots &\vdots &\vdots &\vdots & \vdots & \vdots & \vdots & \vdots\\
V173 & I & 2456435.88824 & 17.745 & 8.717 & 0.011 & 3251.773 & 1764.129 & +39.643 & 192.483 & 5.9879 \\
V173 & I & 2456436.89695 & 17.770 & 8.743 & 0.017 & 3251.773 & 1764.129 & -400.238 & 286.082 & 5.8368 \\
\vdots  & \vdots &      \vdots   & \vdots &\vdots &\vdots &\vdots     &  \vdots     &   \vdots     &  \vdots  & \vdots\\

\hline
\end{tabular}
\end{table*}

\subsection{Astrometry and finding chart}\label{ssec:astro_chart}

To create a reference image with the astrometric information for each star in the field 
covered in NGC~6715, we used the celestial coordinates available in the ACS Globular 
Cluster Survey\footnote{http://www.astro.ufl.edu/\textasciitilde ata/public\_hstgc/} 
\citep[see][]{anderson08+12} which were uploaded for the field of the cluster through GAIA 
(Graphical Astronomy and Image Analysis Tool; \citealt{gaia}). An (x,y) shift to 
match their respective stars in our reference image was applied. Stars lying outside the 
field of view and those without a clear match were removed, and the (x, y) shift was 
refined by minimising the squared coordinate residuals. A total of 305 stars over the 
entire field was used to guarantee that the astrometric solution applied to the reference 
image considered enough stars. The radial Root Mean Square (RMS) scatter obtained in the 
residuals was $\sim$0$^{\prime\prime}$.028 ($\sim$ 0.3 pixels). This astrometrically 
calibrated reference image was used to produce a finding chart for NGC~6715 on which we 
marked the positions and identifications of all variable stars studied in this work (Fig. 
\ref{fig:finding_chart_NGC6715}). Finally, a table with the equatorial J2000 celestial 
coordinates of all variables is given in Table \ref{tab:NGC6715_ephemerides}.

\section{Photometric calibration}\label{sec:calib}

The photometric transformation of instrumental magnitudes to the standard system was 
accomplished using information available in the ACS Globular Cluster Survey, which 
provides calibrated magnitudes for selected stars in the fields of 50 globular clusters 
extracted from images taken with the \textit{Hubble} Space Telescope (HST) instruments ACS 
and WFPC.

\begin{figure}[htp!]
\centering
\includegraphics[scale=1]{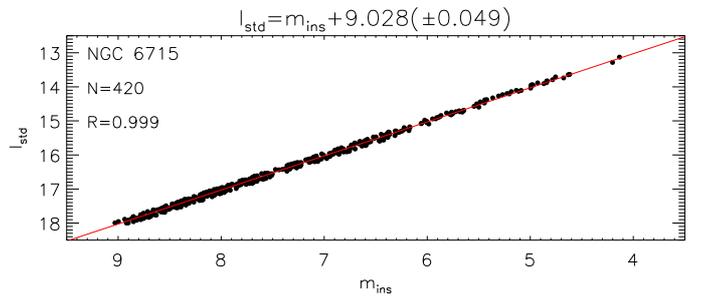}
\caption{Standard I magnitude taken from the HST observations as a function of the instrumental $\mathrm{i}^{\prime}+\mathrm{z}^{\prime}$ magnitude. The red line is the fit that best matches the data and it is described by the equation in the title. The correlation coefficient is 0.999.}
\label{fig:std_calibration}
\end{figure}

By matching the positions of the stars in the field of the HST images with those in our 
reference image, we obtained the photometric transformation shown in Figure 
\ref{fig:std_calibration}. The I magnitude obtained from the ACS 
\citep[see][]{sirianni05+13} is plotted versus the instrumental 
$\mathrm{i}^{\prime}+\mathrm{z}^{\prime}$ magnitude obtained in this study. The red line 
is a linear fit with slope unity yielding the zero point labelled in the title where $N$ 
is the number of stars used in the fit and $R$ is the correlation coefficient obtained.
Due to the substantial differences between the 
$\mathrm{i}^{\prime}+\mathrm{z}^{\prime}$
and standard I wavebands, there are non-linear colour terms in the transformation
that we have not accounted for. However, we have opted for an approximate absolute photometric
calibration since variable star discovery and classification do not require a precise
calibration. Furthermore, our non-standard $\mathrm{i}^{\prime}+\mathrm{z}^{\prime}$
waveband precludes the possibility of using our RR Lyrae light curves for physical
parameter estimation.

\section{Variable star searches}\label{sec:var_search}

Light curves for a total of 1405 stars were obtained with the DanDIA pipeline in the field covered by the reference image. To detect and extract the variable stars from all the non-variables, three automatic (or semi-automatic) techniques were employed. They are described in Sections \ref{sec:rms}, \ref{sec:sb}, and \ref{sec:sum_diff}.

\subsection{Root mean square}\label{sec:rms}

A diagram of root mean square (RMS) magnitude deviation against mean I magnitude (see top Fig. \ref{fig:rms_sb_NGC6715}) was constructed for the cluster. In this diagram, we measure not 
only the photometric scatter for each star, but also the intrinsic variation of the variable stars over time, which gives them a higher RMS than the non-variables. The classification is 
indicated by the colour as detailed in Table \ref{tab:var_type}. To select 
candidate variable stars, we fit a polynomial to the rms values as a function of magnitude 
and flag all
stars with an rms greater than two times the model value.

\subsection{$S_B$ statistic}\label{sec:sb}

A detailed discussion can be found about the benefits of using the $S_B$ statistic to detect variable stars \citep{figuera13+04} and RR Lyrae with Blazhko effect \citep{arellano12+04}. The $S_B$ statistic is defined as 
\begin{equation}\label{eq:sb_index}
S_B=\left(\frac{1}{NM}\right)\sum_{i=1}^M\left(\frac{r_{i,1}}{\sigma_{i,1}}
+\frac{r_{i,2}}{\sigma_{i,2}}+...+\frac{r_{i,k_i}}{\sigma_{i,k_i}}\right)^2,
\end{equation}
where $N$ is the number of data points for a given light curve and $M$ is the number of groups formed of time-consecutive residuals of the same sign from a constant-brightness light curve model (e. g. mean or median). The residuals $r_{i,1}$ to $r_{i,k_i}$ correspond to the $i$th group of $k_i$ time-consecutive residuals of the same sign with corresponding uncertainties $\sigma_{i,1}$ to $\sigma_{i,k_i}$. The $S_B$ statistic is larger in value for light curves with long runs of consecutive data points above or below the mean, which is the case for variable stars with periods longer than the typical photometric cadence.

\begin{figure}[ht]
\centering
\includegraphics[scale=1.0]{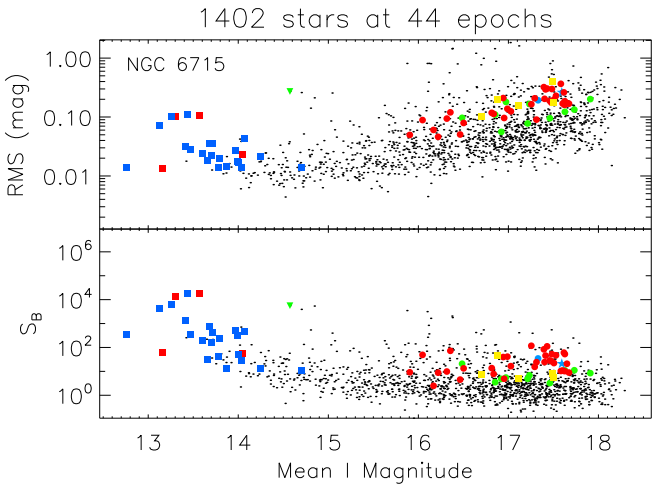}
\caption{Root mean square (RMS) magnitude deviation (top) and $S_B$ statistic (bottom) versus the mean $I$ magnitude for the 1402 stars detected in the field of view of the reference image for NGC~6715. Coloured points follow the convention adopted in Table \ref{tab:var_type} to identify the types of variables found in the field of this globular cluster.}
\label{fig:rms_sb_NGC6715}
\end{figure}

A plot of $S_B$ versus mean I magnitude is given for NGC~6715 (see bottom Fig. 
\ref{fig:rms_sb_NGC6715}), variable stars are plotted in colour. To select candidate 
variable stars,
the same technique employed in Section~\ref{sec:rms} was used, with the threshold also set 
at two times the model $S_B$ values.

\subsection{Stacked difference image}\label{sec:sum_diff}

Based on the results obtained using the DanDIA pipeline, a stacked difference image was 
built for NGC~6715 with the aim of detecting the difference fluxes that correspond to 
variable stars in the field of the reference image. The stacked image is the result of 
summing the absolute values of the difference images divided by the respective pixel 
uncertainty
\begin{equation}\label{eq:sum_diff}
S_{ij}=\sum_k\frac{|D_{kij}|}{\sigma_{kij}},
\end{equation}
where $S_{ij}$ is the stacked image, $D_{kij}$ is the $k$th difference image, 
$\sigma_{kij}$ is the pixel uncertainty associated with each image $k$ and the indexes $i$ 
and $j$ correspond to pixel positions.

All of the variable star candidates obtained by using the RMS and $S_B$ diagrams 
explained in Sections \ref{sec:rms} and \ref{sec:sb} were inspected visually in the 
stacked image and by blinking the difference images to confirm or refute their 
variability.

\section{Variable star classification}\label{sec:var_class}

To define the type of variation in each of the variable stars found, several steps were done. First, we used their position in the colour-magnitude diagram (CMD; Fig. \ref{fig:cmd_NGC6715}) as a reference for their evolutionary stage as most types of variable stars are placed in very well defined zones in the CMD. Second, we implemented a period search for each of the light curves by using the string method \citep{SQmethod} and by minimising the $\chi^2$ in a Fourier analysis fit. Periods found and light curve shapes were also taken into account. Finally, to classify the variable stars, we used the conventions defined in the General Catalogue of Variable Stars \citep{gcvs}.

In Table \ref{tab:var_type}, the classification, corresponding symbols, and colours used in the plots throughout the paper are shown. 

\section{Colour magnitude diagram}\label{sec:cmd}

\begin{figure}[ht]
\centering
\includegraphics[scale=1]{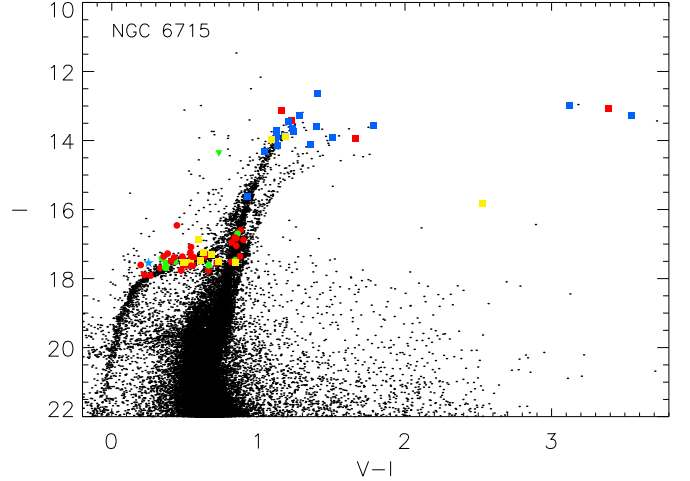}
\caption{Colour magnitude diagram of the globular cluster NGC~6715 built with V and I magnitudes available in the ACS globular cluster survey extracted from HST images.}
\label{fig:cmd_NGC6715}
\end{figure}

As our sample has data available for only one filter, we decided to build the colour-magnitude diagram (CMD; see Fig. \ref{fig:cmd_NGC6715}) by using the information available from the HST images at the ACS Globular Cluster Survey. The data used correspond to the V and I photometry obtained in \cite{sirianni05+13}. The CMD was useful in classifying the variable stars, especially those with poorly defined light curves such as long period variables and semi-regular variables, as well as corroborating cluster membership.

\section{NGC 6715 / C1851-305 / Messier 54}\label{sec:cluster_results}

The details of all variable stars in our FoV that are discussed in this section are listed in Table \ref{tab:NGC6715_ephemerides}, and all light curves are plotted in Figure \ref{fig:lc_variables}.

\begin{table}[htp!]
\caption{Convention used in the variable star classification of this work based on the definitions of the General Catalogue of Variable Stars \citep{gcvs}.}
\label{tab:var_type}
\centering
\begin{tabular}{cccc}
\hline\hline
Type & Id & Point style & Color \\
\hline
\multicolumn{4}{c}{\textbf{Pulsating Variables}} \\
\hline
RR Lyrae        & RR0    & Filled circle               & Red     \\
                & RR1    & Filled circle               & Green   \\
                & RR01   & Filled circle               & Blue    \\
W Virginis      & CWA    & Inverted triangle        & Green    \\
\hline
Semi-regular    & SR     & Filled square            & Red     \\
Long-period irregular    & L      & Filled square               & Blue  \\
\hline
\multicolumn{4}{c}{\textbf{Eclipsing Binaries}} \\
\hline
In general  & E     & Five pointed star   & Blue  \\
\hline
\multicolumn{4}{c}{\textbf{Unclassified Variables}} \\
\hline
In general  & NC     & Filled square               & Yellow  \\

\hline
\end{tabular}
\end{table}

\subsection{Known variables}

This globular cluster has of the order of 200 known variable stars listed in the 
Catalogue of Variable Stars in Galactic Globular Clusters 
\citep[\textbf{CVSGGC;}][]{clement01}. Most of them  are of the RR Lyrae type but a few 
are long-period irregular, semi-regular, W Virginis, eclipsing binaries, and SX Phoenicis. 
To date four studies report variable star discoveries in this globular cluster: V1-V28 
from \cite{rosino52}, V29-V82 from \cite{rosino59+01}, V83-V117 from \cite{layden00+01}, 
and V118-V211 from \cite{sollima10+03}.

\begin{figure*}[ht]
\centering
\includegraphics[scale=1]{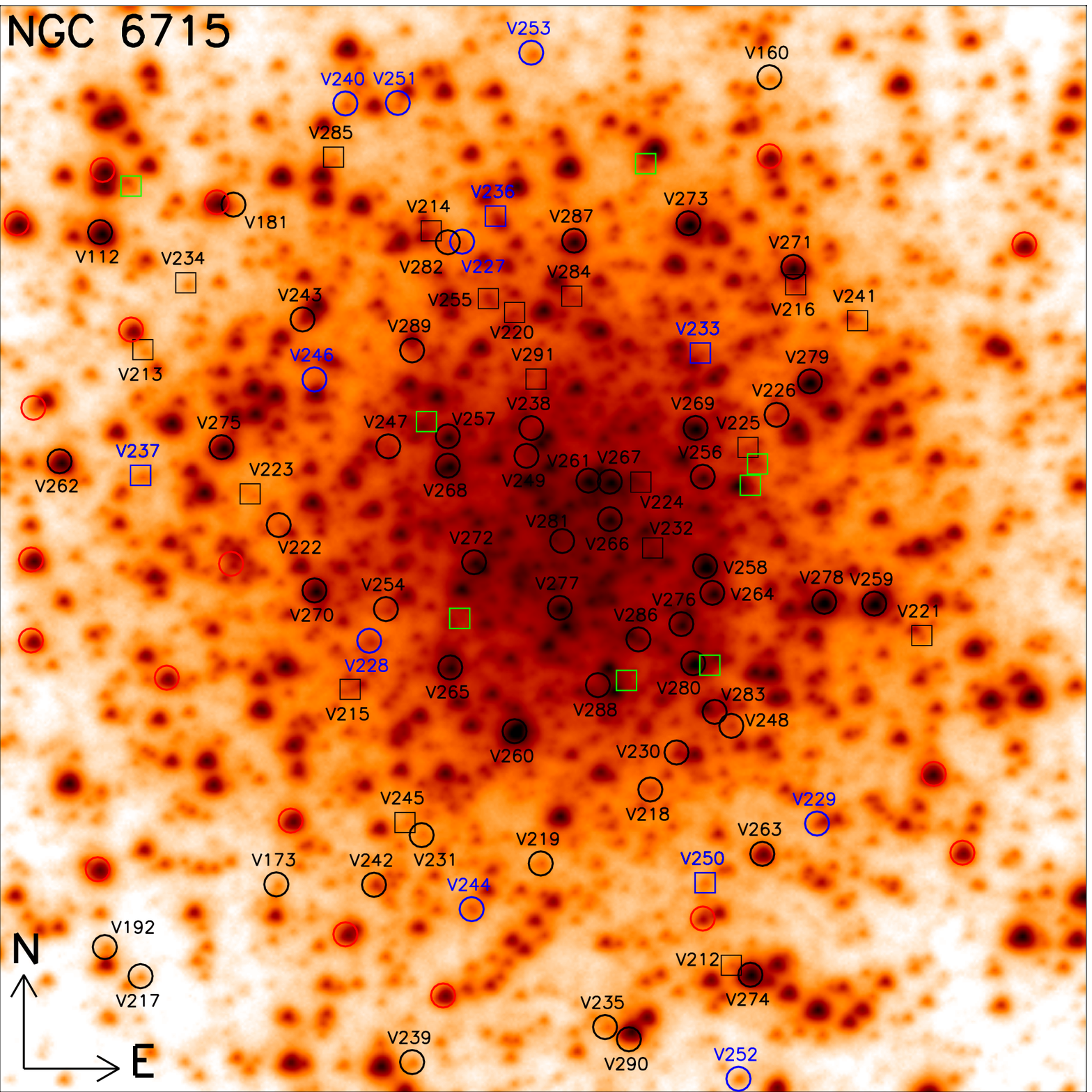}
\caption{Finding chart for the globular cluster NGC~6715. The image used
corresponds to the reference image constructed during the reduction. All known
variables and new discoveries are labelled with their V numbers. Known variables, or new variables discovered in this work, are plotted with black symbols. Variables discovered by the OGLE survey are plotted with blue symbols. Those variables that were candidate variables from \cite{montiel10+01} are plotted using squares. Otherwise symbols are circles. Green squares and red circles, both without labels, are candidate variables from \cite{montiel10+01} and \cite{McDonald14+05} respectively for which we do not detect variability in our survey. Image size is $\sim40\times40$arcsec$^2$.}
\label{fig:finding_chart_NGC6715}
\end{figure*}

In the field of view covered by our reference image there are only five known variable 
stars (V112, V160, V173, V181, V192). All of them lie towards the edges of the image as 
can be seen in Figure \ref{fig:finding_chart_NGC6715}. The star V112 was previously 
classified as long-period irregular. However, we were able to find a period of $\sim$ 100 
days in the variability of this star and due to this we have reclassified it as a 
semi-regular variable. For V160, we were not able to produce a good phased light curve 
using the published period of 0.6194848~d. The discovery observations by 
\cite{sollima10+03} cover a time baseline of only 6 days. With our time baseline of more 
than two years, our derived periods are much more precise and the period found is in 
agreement with that found by the Optical Gravitational Lensing Experiment \citep[OGLE, 
][see below]{udalski92+04}. For V160, we list the OGLE period of 0.62813716~d in 
Table~\ref{tab:NGC6715_ephemerides}. For V173, we improved the period estimate over that 
from \cite{sollima10+03}. For V181, we find a very different period with respect to the 
one estimated by \cite{sollima10+03}. The new period of 0.877072~d makes this RR Lyrae the 
one with the longest period in the cluster. The phased light curve in Figure 
\ref{fig:lc_variables} is somewhat noisy because this variable is highly blended with a 
brighter star. The case of V192 is particularly interesting because the star was 
classified as RR1 with a period of 0.3986799~d. However, at the astrometric position 
reported for this star we found a RR Lyrae type RR0 with a period of 0.600373~d. This is 
also in agreement with the period and classification found by OGLE (see below) which we 
list in Table~\ref{tab:NGC6715_ephemerides}. Again, the phased light curve of this 
variable is noisy because of blending with a brighter star. It is clear then that the 
\cite{sollima10+03} periods and RR Lyrae classifications are not robust based on 
relatively few observations. We will discuss the consequences of this later on.

Recently, \cite{montiel10+01} announced 50 RR Lyrae candidates based on observations 
taken with the HST. However, the data obtained consist of 12 epochs covering only 8 hours, 
which made the study unsuitable for a period search and certainly some RR Lyrae stars will 
have been missed with such a short time baseline. No light curves were presented in their 
paper. Of these 50 candidate variable stars, 17 lie outside of our field of view (VC2, 
VC11-VC18, VC22, VC24, VC38, VC44, VC45, VC47-VC49). As pointed out in the Catalogue of 
Variable Stars in Galactic Globular Clusters \citep{clement01}, 11 of these candidates are 
previously known variables, thus VC2=V127, VC11=V162, VC12=V163, VC13=V95, VC14=V164, 
VC15=V142, VC17=V129, VC18=V179, VC44=V46, VC45=V148, and VC47=V76 (see also 
Appendix~\ref{apen:cross_id}). For 8 of their candidates in our field of view, we could 
not detect variations in our difference images at their coordinates (VC3, VC6, VC19, VC21, 
VC29, VC42, VC43, VC50), and we therefore cannot confirm their variability. We plot their 
positions in Figure \ref{fig:finding_chart_NGC6715} with a green square\footnote{The 
\cite{montiel10+01} coordinates differ by RA$\sim$0.3$^{\prime\prime}$ and 
Dec$\sim$0.8$^{\prime\prime}$ from our coordinates, and we have corrected for this in 
Figure \ref{fig:finding_chart_NGC6715}}. Three of their candidates in our field of view, 
VC28, VC34 and VC46, are the known variables V181, V160 and V192, respectively. We confirm 
the variable nature of the remaining 22 candidates in our field of view, and we have 
assigned them ``V'' numbers as part of our study (see Section \ref{sub:sec:new_var}). 
Table~\ref{tab:NGC6715_ephemerides} lists their VC identification in Column 2. We classify 
18 of them as RR Lyrae stars, one as an eclipsing binary, and are unable to classify VC27, 
VC32, and VC33. We plot their positions in Figure \ref{fig:finding_chart_NGC6715} with a 
square symbol.

The study by \cite{McDonald14+05} using the \textit{VISTA} survey also covered the 
cluster and presents candidate variables based on typically $\sim$12-13 epochs spread over 
$\sim$100-200 days, which was insufficient to derive periods for many of them. Of the 
short period candidate variables listed in their Table 1, 18 are inside the field of view 
covered by our reference image. None of the bright long-period variables listed in their 
Table 2 and faint candidates listed in their Table 3 are inside the field covered in our 
study. Positions of these stars inside our field of view are plotted in Figure 
\ref{fig:finding_chart_NGC6715} with a red circle. It is worth noting that all 
of these candidates are located more toward the edges of the reference image. We detect 
variability in only two of the \cite{McDonald14+05} candidates within our field of view 
(SPVSgr18550405-3028580 and SPVSgr18550386-3028593, which we assign V identifications as 
V229 and V263, respectively -see Section \ref{sub:sec:new_var}).

Similarly, the field of this cluster was also covered by OGLE, particularly with their 
OGLE-IV survey \citep{udalski15+02}. We found in this survey 15 of the variable stars 
studied in our work of which two are previously known variables (V160 and V192) and 13 are 
new discoveries by OGLE (we assigned the following V identifications: 
V227-V229, V233, V236, V237, V240, V244, V246, V250-V253). The OGLE light curves for these stars have typically $\sim$150 epochs covering a baseline of $\sim$2.5 years and OGLE derived precise periods for them. With our data we were also able to find the same periods and type of classification assigned by OGLE. The positions of these stars are plotted in Figure \ref{fig:finding_chart_NGC6715} with a blue colour. Again, it is worth noticing that all of these variables are located more toward the edges of the reference image. In the particular case of V229, V244, and V246, the pipeline was not able to detect these stars in the reference image but their variation is clear in the difference images. Their differential fluxes against phase are plotted in Figure \ref{fig:lc_variables}. Finally, epochs, periods, mean magnitudes, amplitudes, and classifications for these three stars were taken from the OGLE database, although the number of data points listed in Table \ref{tab:NGC6715_ephemerides} correspond to our light curves.

In Appendix~\ref{apen:cross_id} (Table~\ref{tab:ogle_montiel_cross_id}), we provide the cross identifications for the previously known RR Lyrae stars in NGC~6715 between the CVSGGC \citep{clement01}, the variable star candidates from \cite{montiel10+01}, and the OGLE RR Lyrae stars \citep{udalski15+02}.

\subsection{New variables}\label{sub:sec:new_var}

After employing the methods described in Section \ref{sec:var_search}, we were able to extract 67 new variable stars in the core of NGC~6715 of which 30 are RR Lyrae, 1 is a W Virginis star (CWA), 21 are long-period irregular, 3 are semi regular, 1 is an eclipsing binary, and 11 remain without classification.

\subsubsection{RR Lyrae}

\textbf{V213-V226, V230-V232, V234-V235, V238-V239, V241-V243, V245, V247-V249, V254-V255:} These 30 newly discovered variable stars are clear RR Lyrae variables. Their positions in the CMD, light curve shapes, periods, and amplitudes corroborate their variability type. We found that 17 are pulsating in the fundamental mode (RR0); 8 are pulsating in the first overtone (RR1); 1 is a double-mode pulsator (RR01) and 4 RR Lyrae remain with an uncetain subtype (3 RR0? and 1 RR1?). As shown in Table \ref{tab:NGC6715_ephemerides}, their periods range from $\sim$0.28 d to $\sim$0.76 d with amplitudes between 0.06 and 1.69 mag.

The periodogram analysis for V221 showed two predominant frequencies typical of double mode RR Lyrae stars, one equivalent to the fundamental period $P_0=$~0.459608 d and one equivalent to the first overtone period $P_1=$~0.343828 d giving a period ratio $P_1/P_0=$~0.748 which falls in the expected ratio range of $\sim0.725$ to $\sim0.748$ for this type of pulsating RR Lyrae stars \citep{netzel15+02, moskalik13, cox83+02}. Further data for a more detailed analysis of this star will be useful to corroborate its pulsational properties.

\subsubsection{W Virginis}

\textbf{V256:} Particularly interesting is the case of this star as its variation (and position in the CMD) do not follow the pattern found for the other variable stars studied and classified in this work. We found a very well phased light curve with a period of $\sim$14.771 d and an amplitude of 0.71 mag. This is the only bright variable star on the blue side of the colour-magnitude diagram far away from the red giant branch. 

The properties found in the variation of this star and its position in the CMD match very well with the W Virginis type of variable star described in \cite{gcvs}, particularly with the subtype CWA which have periods longer than 8 days \citep[see also][]{wallerstein02}. Although these types of stars have not been commonly found in globular clusters in contrast to RR Lyrae stars, they are not entirely uncommon. In the statistics of variable stars in Galactic globular clusters reported by \cite{clement01} it is possible to notice that 60 variable stars are Cepheids, which include Population II Cepheids, anomalous Cepheids, and RV Tauri stars. V256 is the first CWA star discovered in this cluster.

\subsubsection{Long-period irregular}

\textbf{V260-V280:} These 21 stars are located at the top of the red giant branch as shown with blue squares in Figure \ref{fig:cmd_NGC6715}. Their amplitudes range from 0.05 to 0.46 mag. We found no clear periods for these stars. Due to this and also their position in the colour-magnitude diagram we classified them as long-period irregular. Light curves for all these variables are found in Figure~\ref{fig:lc_variables}.

\subsubsection{Semi regular}

\textbf{V257-V259:} Based on the position of these 3 stars in the colour-magnitude diagram (see Fig.~\ref{fig:cmd_NGC6715}), their periods, and the shape of their light curves, we have classified them as semi-regular. Light curves for these variables may be found in Figure~\ref{fig:lc_variables}. Their amplitudes range from $\sim$0.04 to $\sim$0.45 mag and their periods span between $\sim$20 and $\sim$150 d.

\subsubsection{Eclipsing Binary}

\textbf{V212:} The light curve variations for this star are very similar to those presented in eclipsing binary systems. We found that the amplitude of the deeper eclipse is of the order of $\sim$0.8 mag and the amplitude of the secondary eclipse is $\sim$0.5 mag. The phased light curve shown in Figure \ref{fig:lc_variables} represented a period of $P=$ 0.202144 d.

\subsubsection{Other variable stars}

\textbf{V281-V285:} These 5 stars are clear variable stars. They show clear variability 
by blinking the difference images. They have amplitudes between 0.45 and 0.94 mag. Several 
attempts to determine periods for these stars were done without success. Due to this their 
light curves in Figure \ref{fig:lc_variables} are plotted against HJD. We note that the 
variable source V281 is ony $\sim$0.24 arcsec from the photometric centre of the cluster 
as measured by \cite{goldsbury10}.

\textbf{V286:} Towards the beginning of the 2013 data, we found that the flux of this 
star was increasing to a maximum at $\sim$2456464.9081 d of about $\sim$39400 ADU/s on the 
flux scale of the reference image, corresponding to a peak magnitude of $\sim$15.04 mag. 
After that, its flux strongly decreased during the rest of the observational campaign. 
During 2014 and 2015, we found that the object seems to be at baseline and it is not 
detected in the original images.

The centre of NGC~6715 is $\sim$4.45 arcseconds from this source. Also, at a distance of 
3.73 arcseconds, there is a X-ray source studied by \cite{wrobel11+02} using 
data from \textit{Chandra} and \textit{Hubble} Space Telescope. Given the relatively 
small astrometric uncertainties in these positions, V286 is not associated with either. 
The nature and classification of this variable will need further studies, and it will 
remain without classification in this work.

\textbf{V287-V291:} These 5 stars were not detected by the pipeline in the reference 
image. Unfortunately, there are no light curves available in the OGLE database for these 
stars, but their variations are clear in the difference images. In Figure 
\ref{fig:lc_variables} their differential fluxes against HJD are plotted. As we were not 
able to produce phased light curves for them, they will remain without classification 
until future studies are done.

\section{Oosterhoff dichotomy}\label{sec:oo_type}

In this work we discovered 30 new RR Lyrae, and OGLE also discovered 17 more (of which 13 
are inside our FoV). After removing the stars without a secure classification, there are 
33 new RR0 and 9 new RR1 stars, which represent a significant increment in the known RR 
Lyrae population. The vast majority of these new RR Lyrae stars are cluster members since 
we have studied the core of NGC 6715. Hence it is pertinent to recalculate the mean 
periods and number ratios of the RR Lyrae stars to see if it modifies the current 
conclusions about the Oosterhoff type of NGC~6715.

\begin{figure}[ht]
\centering
\includegraphics[scale=1]{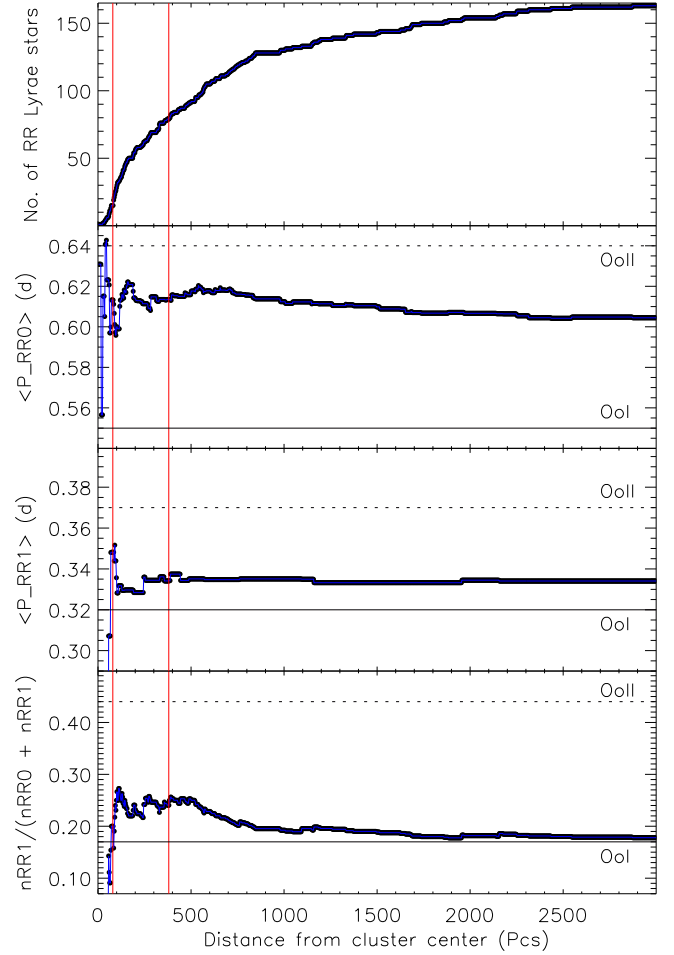}
\caption{The number of RR Lyraes, mean periods, and number ratios as a function of the distance from the cluster centre. The red lines correspond to the tidal radii calculated by \cite{mcLaughlin05+01}.}
\label{fig:rr_from center}
\end{figure}

However, we note that NGC~6715 lies projected against the Sagittarius Dwarf Spheroidal 
galaxy and behind the Galactic bulge. Therefore, the sample of RR Lyrae stars in the field 
of the cluster is a mixture of cluster members, Bulge stars, and Sgr dSph stars. Towards 
the centre of NGC~6715, the cluster member RR Lyraes dominate, but limiting our sample to 
the cluster core also reduces the number of RR Lyraes that can be used to calculate the 
mean periods and number ratios. Hence, in Figure \ref{fig:rr_from center}, we have plotted 
the total number of RR Lyrae stars, the mean period of the RR0 stars, the mean period of 
the RR1 stars, and the number ratio of the RR1 to all RR Lyrae stars (nRR1/(nRR0 + nRR1)), 
all as a function of distance from the cluster centre \citep[RA(J2000) = 18:55:03.33; 
Dec(J2000)=$-$30:28:47.5;][]{goldsbury10}. To convert angles on the sky into distances in 
parsecs we used the distance from the Sun to NGC~6715 of 26500 pc \citep[][2010 
version]{harriscatalog}. The inner and outer red vertical lines are the tidal radii 
estimated by \cite{mcLaughlin05+01} based on \cite{king66} and \cite{wilson75} models, 
respectively. The horizontal black lines correspond to the OoI (solid) and OoII (dashed) 
types given by \cite[][]{smith95} in his Table 3.2. We only used RR Lyrae stars with 
certain classifications (i.e. published phased light curves with reliable period 
estimates).

Figure~\ref{fig:rr_from center} clearly shows that within the range of the two estimates 
of the tidal radii, where the RR Lyrae stars from the cluster still dominate, the values 
of $<P_{\mathrm{RR0}}>$, $<P_{\mathrm{RR1}}>$ and nRR1/(nRR0 + nRR1) are intermediate 
between those expected for OoI and OoII clusters. We obtain $<P_{\mathrm{RR0}}>$=0.613529 
d, $<P_{\mathrm{RR1}}>$=0.334107 d, and nRR1/(nRR0 + nRR1)= 0.24050633 when calculated for 
the 79 RR Lyraes within the outer estimate of the tidal radius (i.e. 380 pc). The 
contaminating RR Lyrae populations have ($<P_{\mathrm{RR0}}>$, 
$<P_{\mathrm{RR1}}>$)=(0.556 d, 0.310 d) and 
($<P_{\mathrm{RR0}}>$,$<P_{\mathrm{RR1}}>$)=(0.574 d, 0.322 d) for the Bulge and Sgr dSph, 
respectively \citep{soszynski11+09, cseresnjes01}, placing them in the OoI region in the 
$<P_{\mathrm{RR0}}>$-metallicity diagram of \cite{catelan09}. This explains why the values 
of $<P_{\mathrm{RR0}}>$ and nRR1/(nRR0 + nRR1) tend towards the OoI values beyond the 
outer estimate of the tidal radius. The value of $<P_{\mathrm{RR1}}>$ is hardly influenced 
though because there are only 6 RR1 stars beyond the outer estimate of the tidal radius. 
Hence the new RR Lyrae discoveries in this paper have served to confirm that NGC~6715 is 
of intermediate Oosterhoff type.

\section{Conclusions}\label{sec:conclusion}

The globular cluster NGC~6715 turns out to be a very interesting stellar system,
for which images with the highest angular resolution have ever been obtained so far with ground-based 
telescopes. The use of the EMCCD and shift-and-add technique was demonstrated to be an
excellent procedure to minimise the effect of the atmospheric turbulence which is one of the main 
constraints when doing ground-based observations. Thanks to this and the use of difference image
analysis it was possible to obtain high-precision time series photometry in the core of this cluster
down to $I\sim$18.3 mag.

A total of 1405 stars in the field covered by the reference image were statistically 
studied for variable star detection.
We presented light curves for 17 previously known variables that were found toward the edges of the reference image (16 RR Lyrae and 1 SR).
We also discovered 67 new variable stars, which consist of 30 RR Lyrae, 21 long-period irregular, 3 semi-regular, 1 W Virginis,
1 eclipsing binary, and 11 unclassified stars. We estimated periods and ephemerides for all variable stars in the field of our reference image.
Our new RR Lyrae star discoveries help confirm that NGC~6715 is of intermediate Oosterhoff type.
Finally, our photometric measurements for all variable stars studied in this work are 
available in electronic form through the Strasbourg Astronomical Data Centre.

\begin{acknowledgements}
Our thanks go to Christine Clement for clarifying the known variable star content in NGC~6715 and the numbering systems of the variable stars while we were working on these clusters. This support to the astronomical community is very much appreciated. The Danish 1.54m telescope is operated based on a grant from the \emph{Danish Natural Science Foundation (FNU)}. This publication was made possible by NPRP grant \# X-019-1-006 from the \emph{Qatar National Research Fund (a member of Qatar Foundation)}. The statements made herein are solely the responsibility of the authors. KH acknowledges support from STFC grant ST/M001296/1. GD acknowledges Regione Campania for support from POR-FSE Campania 2014-2020. D.F.E. is funded by the UK Science and Technology Facilities Council. TH is supported by a Sapere Aude Starting Grant from the Danish Council for Independent Research. Research at Centre for Star and Planet Formation is funded by the Danish National Research Foundation. TCH acknowledges support from the Korea Research Council of Fundamental Science \& Technology (KRCF) via the KRCF Young Scientist Research Fellowship.  Programme and for financial support from KASI travel grant number 2013-9-400-00, 2014-1-400-06 \& 2015-1-850-04. NP acknowledges funding by the Gemini-Conicyt Fund, allocated to project No. 32120036 and by the Portuguese FCT - Foundation for Science and Technology and the European Social Fund (ref: SFRH/BGCT/113686/2015). CITEUC is funded by National Funds through FCT - Foundation for Science and Technology (project: UID/Multi/00611/2013) and FEDER - European Regional Development Fund through
COMPETE 2020 -Operational Programme Competitiveness and Internationalisation (project: POCI-01-0145-FEDER-006922). OW and J. Surdej acknowledge support from the Communaut\'{e} fran\c{c}aise de Belgique - Actions de recherche concert\'{e}es - Acad\'{e}mie Wallonie-Europe. This work has made extensive use of the \emph{ADS} and \emph{SIMBAD} services, for which we are thankful.

\end{acknowledgements}
\bibliographystyle{biblio}

\begin{thebibliography}{64}
\expandafter\ifx\csname natexlab\endcsname\relax\def\natexlab#1{#1}\fi

\bibitem[{{Alard}(2000)}]{alard00}
{Alard}, C. 2000, \aaps, 144, 363

\bibitem[{{Alard} \& {Lupton}(1998)}]{alard98+01}
{Alard}, C. \& {Lupton}, R.~H. 1998, \apj, 503, 325

\bibitem[{{Anderson} {et~al.}(2008){Anderson}, {Sarajedini}, {Bedin}, {King},
  {Piotto}, {Reid}, {Siegel}, {Majewski}, {Paust}, {Aparicio}, {Milone},
  {Chaboyer}, \& {Rosenberg}}]{anderson08+12}
{Anderson}, J., {Sarajedini}, A., {Bedin}, L.~R., {et~al.} 2008, \aj, 135, 2055

\bibitem[{{Arellano Ferro} {et~al.}(2012){Arellano Ferro}, {Bramich}, {Figuera
  Jaimes}, {Giridhar}, \& {Kuppuswamy}}]{arellano12+04}
{Arellano Ferro}, A., {Bramich}, D.~M., {Figuera Jaimes}, R., {Giridhar}, S.,
  \& {Kuppuswamy}, K. 2012, MNRAS, 420, 1333

\bibitem[{{Bellazzini} {et~al.}(2008){Bellazzini}, {Ibata}, {Chapman},
  {Mackey}, {Monaco}, {Irwin}, {Martin}, {Lewis}, \&
  {Dalessandro}}]{bellazzini08+08}
{Bellazzini}, M., {Ibata}, R.~A., {Chapman}, S.~C., {et~al.} 2008, \aj, 136,
  1147

\bibitem[{{Bessell}(2005)}]{bessell05}
{Bessell}, M.~S. 2005, ARAA, 43, 293

\bibitem[{{Bozza} {et~al.}(2016){Bozza}, {Shvartzvald}, {Udalski}, {Calchi
  Novati}, {Bond}, {Han}, {Hundertmark}, {Poleski}, {Pawlak}, {Szyma{\'n}ski},
  {Skowron}, {Mr{\'o}z}, {Koz{\l}owski}, {Wyrzykowski}, {Pietrukowicz},
  {Soszy{\'n}ski}, {Ulaczyk}, {OGLE Group}, {and}, {Beichman}, {Bryden},
  {Carey}, {Fausnaugh}, {Gaudi}, {Gould}, {Henderson}, {Pogge}, {Wibking},
  {Yee}, {Zhu}, {Spitzer Team}, {Abe}, {Asakura}, {Barry}, {Bennett},
  {Bhattacharya}, {Donachie}, {Freeman}, {Fukui}, {Hirao}, {Inayama}, {Itow},
  {Koshimoto}, {Li}, {Ling}, {Masuda}, {Matsubara}, {Muraki}, {Nagakane},
  {Nishioka}, {Ohnishi}, {Oyokawa}, {Rattenbury}, {Saito}, {Sharan},
  {Sullivan}, {Sumi}, {Suzuki}, {Tristram}, {Wakiyama}, {Yonehara}, {MOA
  Group}, {Choi}, {Park}, {Jung}, {Shin}, {Albrow}, {Park}, {Kim}, {Lee},
  {Cha}, {Kim}, {Lee}, {KMTNet Group}, {Dominik}, {J{\o}rgensen}, {Andersen},
  {Bramich}, {Burgdorf}, {Ciceri}, {D'Ago}, {Evans}, {Figuera Jaimes}, {Gu},
  {Hinse}, {Kains}, {Kerins}, {Korhonen}, {Kuffmeier}, {Mancini}, {Popovas},
  {Rabus}, {Rahvar}, {Rasmussen}, {Scarpetta}, {Skottfelt}, {Snodgrass},
  {Southworth}, {Surdej}, {Unda-Sanzana}, {von Essen}, {Wang}, {Wertz},
  {MiNDSTEp}, {Maoz}, {Friedmann}, {Kaspi}, \& {Wise Group}}]{bozza16}
{Bozza}, V., {Shvartzvald}, Y., {Udalski}, A., {et~al.} 2016, \apj, 820, 79

\bibitem[{{Bramich}(2008)}]{bramich08}
{Bramich}, D.~M. 2008, MNRAS, 386, L77

\bibitem[{{Bramich} {et~al.}(2013){Bramich}, {Horne}, {Albrow}, {Tsapras},
  {Snodgrass}, {Street}, {Hundertmark}, {Kains}, {Arellano Ferro}, {Figuera},
  \& {Giridhar}}]{bramich13+10}
{Bramich}, D.~M., {Horne}, K., {Albrow}, M.~D., {et~al.} 2013, MNRAS, 428, 2275

\bibitem[{{Carretta} {et~al.}(2010){Carretta}, {Bragaglia}, {Gratton},
  {Lucatello}, {Bellazzini}, {Catanzaro}, {Leone}, {Momany}, {Piotto}, \&
  {D'Orazi}}]{carretta10+09}
{Carretta}, E., {Bragaglia}, A., {Gratton}, R.~G., {et~al.} 2010, \aap, 520,
  A95

\bibitem[{{Catelan}(2004)}]{catelan04}
{Catelan}, M. 2004, in Astronomical Society of the Pacific Conference Series,
  Vol. 310, IAU Colloq. 193: Variable Stars in the Local Group, ed. D.~W.
  {Kurtz} \& K.~R. {Pollard}, 113

\bibitem[{{Catelan}(2009)}]{catelan09}
{Catelan}, M. 2009, \apss, 320, 261

\bibitem[{{Ciceri} {et~al.}(2016){Ciceri}, {Mancini}, {Southworth}, {Lendl},
  {Tregloan-Reed}, {Brahm}, {Chen}, {D'Ago}, {Dominik}, {Figuera Jaimes},
  {Galianni}, {Harps{\o}e}, {Hinse}, {J{\o}rgensen}, {Juncher}, {Korhonen},
  {Liebig}, {Rabus}, {Bonomo}, {Bott}, {Henning}, {Jord{\'a}n}, {Sozzetti},
  {Alsubai}, {Andersen}, {Bajek}, {Bozza}, {Bramich}, {Browne}, {Calchi
  Novati}, {Damerdji}, {Diehl}, {Elyiv}, {Giannini}, {Gu}, {Hundertmark},
  {Kains}, {Penny}, {Popovas}, {Rahvar}, {Scarpetta}, {Schmidt}, {Skottfelt},
  {Snodgrass}, {Surdej}, {Vilela}, {Wang}, \& {Wertz}}]{ciceri16+mindstep}
{Ciceri}, S., {Mancini}, L., {Southworth}, J., {et~al.} 2016, \mnras, 456, 990

\bibitem[{{Clement} {et~al.}(2001){Clement}, {Muzzin}, {Dufton}, {Ponnampalam},
  {Wang}, {Burford}, {Richardson}, {Rosebery}, {Rowe}, \& {Hogg}}]{clement01}
{Clement}, C.~M., {Muzzin}, A., {Dufton}, Q., {et~al.} 2001, AJ, 122, 2587

\bibitem[{{Cox} {et~al.}(1983){Cox}, {Hodson}, \& {Clancy}}]{cox83+02}
{Cox}, A.~N., {Hodson}, S.~W., \& {Clancy}, S.~P. 1983, \apj, 266, 94

\bibitem[{{Cseresnjes}(2001)}]{cseresnjes01}
{Cseresnjes}, P. 2001, \aap, 375, 909

\bibitem[{{Draper}(2000)}]{gaia}
{Draper}, P.~W. 2000, in Astronomical Society of the Pacific Conference Series,
  Vol. 216, Astronomical Data Analysis Software and Systems IX, ed.
  N.~{Manset}, C.~{Veillet}, \& D.~{Crabtree}, 615

\bibitem[{{Evans} {et~al.}(2016){Evans}, {Southworth}, {Maxted}, {Skottfelt},
  {Hundertmark}, {J{\o}rgensen}, {Dominik}, {Alsubai}, {Andersen}, {Bozza},
  {Bramich}, {Burgdorf}, {Ciceri}, {D'Ago}, {Figuera Jaimes}, {Gu},
  {Haugb{\o}lle}, {Hinse}, {Juncher}, {Kains}, {Kerins}, {Korhonen},
  {Kuffmeier}, {Mancini}, {Peixinho}, {Popovas}, {Rabus}, {Rahvar}, {Schmidt},
  {Snodgrass}, {Starkey}, {Surdej}, {Tronsgaard}, {von Essen}, {Wang}, \&
  {Wertz}}]{evans16+mindstep}
{Evans}, D.~F., {Southworth}, J., {Maxted}, P.~F.~L., {et~al.} 2016, \aap, 589,
  A58

\bibitem[{{Figuera Jaimes} {et~al.}(2013){Figuera Jaimes}, {Arellano Ferro},
  {Bramich}, {Giridhar}, \& {Kuppuswamy}}]{figuera13+04}
{Figuera Jaimes}, R., {Arellano Ferro}, A., {Bramich}, D.~M., {Giridhar}, S.,
  \& {Kuppuswamy}, K. 2013, \aap, 556, A20

\bibitem[{{Figuera Jaimes} {et~al.}(2016){Figuera Jaimes}, {Bramich},
  {Skottfelt}, {Kains}, {J{\o}rgensen}, {Horne}, {Dominik}, {Alsubai}, {Bozza},
  {Calchi Novati}, {Ciceri}, {D'Ago}, {Galianni}, {Gu}, {Harps{\o}e},
  {Haugb{\o}lle}, {Hinse}, {Hundertmark}, {Juncher}, {Korhonen}, {Mancini},
  {Popovas}, {Rabus}, {Rahvar}, {Scarpetta}, {Schmidt}, {Snodgrass},
  {Southworth}, {Starkey}, {Street}, {Surdej}, {Wang}, \& {Wertz}}]{figuera15}
{Figuera Jaimes}, R., {Bramich}, D.~M., {Skottfelt}, J., {et~al.} 2016, \aap,
  588, A128

\bibitem[{{Goldsbury} {et~al.}(2010){Goldsbury}, {Richer}, {Anderson},
  {Dotter}, {Sarajedini}, \& {Woodley}}]{goldsbury10}
{Goldsbury}, R., {Richer}, H.~B., {Anderson}, J., {et~al.} 2010, \aj, 140, 1830

\bibitem[{{Harps{\o}e} {et~al.}(2012){Harps{\o}e}, {J{\o}rgensen}, {Andersen},
  \& {Grundahl}}]{harpsoe12+03}
{Harps{\o}e}, K.~B.~W., {J{\o}rgensen}, U.~G., {Andersen}, M.~I., \&
  {Grundahl}, F. 2012, \aap, 542, A23

\bibitem[{{Harris}(1996)}]{harriscatalog}
{Harris}, W.~E. 1996, AJ, 112, 1487

\bibitem[{{Ibata} {et~al.}(2009){Ibata}, {Bellazzini}, {Chapman},
  {Dalessandro}, {Ferraro}, {Irwin}, {Lanzoni}, {Lewis}, {Mackey}, {Miocchi},
  \& {Varghese}}]{ibata09+10}
{Ibata}, R., {Bellazzini}, M., {Chapman}, S.~C., {et~al.} 2009, \apjl, 699,
  L169

\bibitem[{{Jerram} {et~al.}(2001){Jerram}, {Pool}, {Bell}, {Burt}, {Bowring},
  {Spencer}, {Hazelwood}, {Moody}, {Catlett}, \& {Heyes}}]{jerram01+09}
{Jerram}, P., {Pool}, P.~J., {Bell}, R., {et~al.} 2001, in Society of
  Photo-Optical Instrumentation Engineers (SPIE) Conference Series, Vol. 4306,
  Sensors and Camera Systems for Scientific, Industrial, and Digital
  Photography Applications II, ed. M.~M. {Blouke}, J.~{Canosa}, \& N.~{Sampat},
  178--186

\bibitem[{{King}(1966)}]{king66}
{King}, I.~R. 1966, \aj, 71, 64

\bibitem[{{Kinman}(1959)}]{kinman59}
{Kinman}, T.~D. 1959, \mnras, 119, 134

\bibitem[{{Lafler} \& {Kinman}(1965)}]{SQmethod}
{Lafler}, J. \& {Kinman}, T.~D. 1965, \apjs, 11, 216

\bibitem[{{Layden} \& {Sarajedini}(2000)}]{layden00+01}
{Layden}, A.~C. \& {Sarajedini}, A. 2000, \aj, 119, 1760

\bibitem[{{Lee} \& {Carney}(1999)}]{lee99+01}
{Lee}, J.-W. \& {Carney}, B.~W. 1999, \aj, 118, 1373

\bibitem[{{Li} \& {Qian}(2013)}]{li13+01}
{Li}, K. \& {Qian}, S.-B. 2013, \na, 22, 57

\bibitem[{{Majewski} {et~al.}(2003){Majewski}, {Skrutskie}, {Weinberg}, \&
  {Ostheimer}}]{majewski03+03}
{Majewski}, S.~R., {Skrutskie}, M.~F., {Weinberg}, M.~D., \& {Ostheimer}, J.~C.
  2003, \apj, 599, 1082

\bibitem[{{Mancini} {et~al.}(2015){Mancini}, {Giacobbe}, {Littlefair},
  {Southworth}, {Bozza}, {Damasso}, {Dominik}, {Hundertmark}, {J{\o}rgensen},
  {Juncher}, {Popovas}, {Rabus}, {Rahvar}, {Schmidt}, {Skottfelt}, {Snodgrass},
  {Sozzetti}, {Alsubai}, {Bramich}, {Calchi Novati}, {Ciceri}, {D'Ago},
  {Figuera Jaimes}, {Galianni}, {Gu}, {Harps{\o}e}, {Haugb{\o}lle}, {Henning},
  {Hinse}, {Kains}, {Korhonen}, {Scarpetta}, {Starkey}, {Surdej}, {Wang}, \&
  {Wertz}}]{mancini15+mindstep}
{Mancini}, L., {Giacobbe}, P., {Littlefair}, S.~P., {et~al.} 2015, \aap, 584,
  A104

\bibitem[{{McDonald} {et~al.}(2014){McDonald}, {Zijlstra}, {Sloan}, {Kerins},
  {Lagadec}, \& {Minniti}}]{McDonald14+05}
{McDonald}, I., {Zijlstra}, A.~A., {Sloan}, G.~C., {et~al.} 2014, \mnras, 439,
  2618

\bibitem[{{McLaughlin} \& {van der Marel}(2005)}]{mcLaughlin05+01}
{McLaughlin}, D.~E. \& {van der Marel}, R.~P. 2005, \apjs, 161, 304

\bibitem[{{Milone}(2015)}]{milone15}
{Milone}, A.~P. 2015, ArXiv e-prints [\eprint[arXiv]{1510.02578}]

\bibitem[{{Monaco} {et~al.}(2005){Monaco}, {Bellazzini}, {Ferraro}, \&
  {Pancino}}]{monaco05+03}
{Monaco}, L., {Bellazzini}, M., {Ferraro}, F.~R., \& {Pancino}, E. 2005,
  \mnras, 356, 1396

\bibitem[{{Montiel} \& {Mighell}(2010)}]{montiel10+01}
{Montiel}, E.~J. \& {Mighell}, K.~J. 2010, \aj, 140, 1500

\bibitem[{{Moskalik}(2013)}]{moskalik13}
{Moskalik}, P. 2013, in Astrophysics and Space Science Proceedings, Vol.~31,
  Stellar Pulsations: Impact of New Instrumentation and New Insights, ed. J.~C.
  {Su{\'a}rez}, R.~{Garrido}, L.~A. {Balona}, \& J.~{Christensen-Dalsgaard},
  103

\bibitem[{{Netzel} {et~al.}(2015){Netzel}, {Smolec}, \&
  {Dziembowski}}]{netzel15+02}
{Netzel}, H., {Smolec}, R., \& {Dziembowski}, W. 2015, \mnras, 451, L25

\bibitem[{{Oosterhoff}(1939)}]{oosterhoff39}
{Oosterhoff}, P.~T. 1939, The Observatory, 62, 104

\bibitem[{{Pryor} \& {Meylan}(1993)}]{pryor93+01}
{Pryor}, C. \& {Meylan}, G. 1993, in Astronomical Society of the Pacific
  Conference Series, Vol.~50, Structure and Dynamics of Globular Clusters, ed.
  S.~G. {Djorgovski} \& G.~{Meylan}, 357

\bibitem[{{Rosenberg} {et~al.}(2004){Rosenberg}, {Recio-Blanco}, \&
  {Garc{\'{\i}}a-Mar{\'{\i}}n}}]{rosenberg04+02}
{Rosenberg}, A., {Recio-Blanco}, A., \& {Garc{\'{\i}}a-Mar{\'{\i}}n}, M. 2004,
  \apj, 603, 135

\bibitem[{{Rosino}(1952)}]{rosino52}
{Rosino}, L. 1952, \memsai, 23, 49

\bibitem[{{Rosino} \& {Nobili}(1958)}]{rosino59+01}
{Rosino}, L. \& {Nobili}, F. 1958, \memsai, 29, 413

\bibitem[{{Samus} {et~al.}(2009){Samus}, {Durlevich}, \& {et al.}}]{gcvs}
{Samus}, N.~N., {Durlevich}, O.~V., \& {et al.} 2009, VizieR Online Data
  Catalog, 1, 2025

\bibitem[{{Siegel} {et~al.}(2007){Siegel}, {Dotter}, {Majewski}, {Sarajedini},
  {Chaboyer}, {Nidever}, {Anderson}, {Mar{\'{\i}}n-Franch}, {Rosenberg},
  {Bedin}, {Aparicio}, {King}, {Piotto}, \& {Reid}}]{siegel07+13}
{Siegel}, M.~H., {Dotter}, A., {Majewski}, S.~R., {et~al.} 2007, \apjl, 667,
  L57

\bibitem[{{Sirianni} {et~al.}(2005){Sirianni}, {Jee}, {Ben{\'{\i}}tez},
  {Blakeslee}, {Martel}, {Meurer}, {Clampin}, {De Marchi}, {Ford}, {Gilliland},
  {Hartig}, {Illingworth}, {Mack}, \& {McCann}}]{sirianni05+13}
{Sirianni}, M., {Jee}, M.~J., {Ben{\'{\i}}tez}, N., {et~al.} 2005, \pasp, 117,
  1049

\bibitem[{{Skottfelt} {et~al.}(2015{\natexlab{a}}){Skottfelt}, {Bramich},
  {Figuera Jaimes}, {J{\o}rgensen}, {Kains}, {Arellano Ferro}, {Alsubai},
  {Bozza}, {Calchi Novati}, {Ciceri}, {D'Ago}, {Dominik}, {Galianni}, {Gu},
  {Harps{\o}e}, {Haugb{\o}lle}, {Hinse}, {Hundertmark}, {Juncher}, {Korhonen},
  {Liebig}, {Mancini}, {Popovas}, {Rabus}, {Rahvar}, {Scarpetta}, {Schmidt},
  {Snodgrass}, {Southworth}, {Starkey}, {Street}, {Surdej}, {Wang}, \& {Wertz
  (The Mindstep Consortium)}}]{Skottfelt15+05}
{Skottfelt}, J., {Bramich}, D.~M., {Figuera Jaimes}, R., {et~al.}
  2015{\natexlab{a}}, \aap, 573, A103

\bibitem[{{Skottfelt} {et~al.}(2013){Skottfelt}, {Bramich}, {Figuera Jaimes},
  {J{\o}rgensen}, {Kains}, {Harps{\o}e}, {Liebig}, {Penny}, {Alsubai},
  {Andersen}, {Bozza}, {Browne}, {Calchi Novati}, {Damerdji}, {Diehl},
  {Dominik}, {Elyiv}, {Giannini}, {Hessman}, {Hinse}, {Hundertmark}, {Juncher},
  {Kerins}, {Korhonen}, {Mancini}, {Martin}, {Rabus}, {Rahvar}, {Scarpetta},
  {Southworth}, {Snodgrass}, {Street}, {Surdej}, {Tregloan-Reed}, {Vilela}, \&
  {Williams}}]{skottfelt13}
{Skottfelt}, J., {Bramich}, D.~M., {Figuera Jaimes}, R., {et~al.} 2013, \aap,
  553, A111

\bibitem[{{Skottfelt} {et~al.}(2015{\natexlab{b}}){Skottfelt}, {Bramich},
  {Hundertmark}, {J{\o}rgensen}, {Michaelsen}, {Kj{\ae}rgaard}, {Southworth},
  {S{\o}rensen}, {Andersen}, {Andersen}, {Christensen-Dalsgaard}, {Frandsen},
  {Grundahl}, {Harps{\o}e}, {Kjeldsen}, \& {Pall{\'e}}}]{Skottfelt15+15}
{Skottfelt}, J., {Bramich}, D.~M., {Hundertmark}, M., {et~al.}
  2015{\natexlab{b}}, \aap, 574, A54

\bibitem[{{Smith}(1995)}]{smith95}
{Smith}, H.~A. 1995, Cambridge Astrophysics Series, 27

\bibitem[{{Smith} {et~al.}(2011){Smith}, {Catelan}, \& {Kuehn}}]{smith11+03}
{Smith}, H.~A., {Catelan}, M., \& {Kuehn}, C. 2011, in RR Lyrae Stars,
  Metal-Poor Stars, and the Galaxy, ed. A.~{McWilliam}, Vol.~5, 17

\bibitem[{{Smith} {et~al.}(2008){Smith}, {Giltinan}, {O'Connor}, {O'Driscoll},
  {Collins}, {Loughnan}, \& {Papageorgiou}}]{smith08+06}
{Smith}, N., {Giltinan}, A., {O'Connor}, A., {et~al.} 2008, in Astrophysics and
  Space Science Library, Vol. 351, Astrophysics and Space Science Library, ed.
  D.~{Phelan}, O.~{Ryan}, \& A.~{Shearer}, 257

\bibitem[{{Sollima} {et~al.}(2010){Sollima}, {Cacciari}, {Bellazzini}, \&
  {Colucci}}]{sollima10+03}
{Sollima}, A., {Cacciari}, C., {Bellazzini}, M., \& {Colucci}, S. 2010, \mnras,
  406, 329

\bibitem[{{Sollima} {et~al.}(2014){Sollima}, {Cassisi}, {Fiorentino}, \&
  {Gratton}}]{sollima14+03}
{Sollima}, A., {Cassisi}, S., {Fiorentino}, G., \& {Gratton}, R.~G. 2014,
  \mnras, 444, 1862

\bibitem[{{Soszy{\'n}ski} {et~al.}(2011){Soszy{\'n}ski}, {Udalski},
  {Pietrukowicz}, {Szyma{\'n}ski}, {Kubiak}, {Pietrzy{\'n}ski}, {Wyrzykowski},
  {Ulaczyk}, {Poleski}, \& {Koz{\l}owski}}]{soszynski11+09}
{Soszy{\'n}ski}, I., {Udalski}, A., {Pietrukowicz}, P., {et~al.} 2011, \actaa,
  61, 285

\bibitem[{{Southworth} {et~al.}(2015){Southworth}, {Mancini}, {Tregloan-Reed},
  {Calchi Novati}, {Ciceri}, {D'Ago}, {Delrez}, {Dominik}, {Evans}, {Gillon},
  {Jehin}, {J{\o}rgensen}, {Haugb{\o}lle}, {Lendl}, {Arena}, {Barbieri},
  {Barbieri}, {Corfini}, {Lopresti}, {Marchini}, {Marino}, {Alsubai}, {Bozza},
  {Bramich}, {Jaimes}, {Hinse}, {Henning}, {Hundertmark}, {Juncher},
  {Korhonen}, {Popovas}, {Rabus}, {Rahvar}, {Schmidt}, {Skottfelt},
  {Snodgrass}, {Starkey}, {Surdej}, \& {Wertz}}]{southworth15+mindstep}
{Southworth}, J., {Mancini}, L., {Tregloan-Reed}, J., {et~al.} 2015, \mnras,
  454, 3094

\bibitem[{{Street} {et~al.}(2016){Street}, {Udalski}, {Calchi Novati},
  {Hundertmark}, {Zhu}, {Gould}, {Yee}, {Tsapras}, {Bennett}, {RoboNet
  Project}, {Consortium}, {J{\o}rgensen}, {Dominik}, {Andersen}, {Bachelet},
  {Bozza}, {Bramich}, {Burgdorf}, {Cassan}, {Ciceri}, {D'Ago}, {Dong}, {Evans},
  {Gu}, {Harkonnen}, {Hinse}, {Horne}, {Figuera Jaimes}, {Kains}, {Kerins},
  {Korhonen}, {Kuffmeier}, {Mancini}, {Menzies}, {Mao}, {Peixinho}, {Popovas},
  {Rabus}, {Rahvar}, {Ranc}, {Tronsgaard Rasmussen}, {Scarpetta}, {Schmidt},
  {Skottfelt}, {Snodgrass}, {Southworth}, {Steele}, {Surdej}, {Unda-Sanzana},
  {Verma}, {von Essen}, {Wambsganss}, {Wang}, {Wertz}, {OGLE Project},
  {Poleski}, {Pawlak}, {Szyma{\'n}ski}, {Skowron}, {Mr{\'o}z}, {Koz{\l}owski},
  {Wyrzykowski}, {Pietrukowicz}, {Pietrzy{\'n}ski}, {Soszy{\'n}ski}, {Ulaczyk},
  {Spitzer Team}, {Beichman}, {Bryden}, {Carey}, {Gaudi}, {Henderson}, {Pogge},
  {Shvartzvald}, {MOA Collaboration}, {Abe}, {Asakura}, {Bhattacharya}, {Bond},
  {Donachie}, {Freeman}, {Fukui}, {Hirao}, {Inayama}, {Itow}, {Koshimoto},
  {Li}, {Ling}, {Masuda}, {Matsubara}, {Muraki}, {Nagakane}, {Nishioka},
  {Ohnishi}, {Oyokawa}, {Rattenbury}, {Saito}, {Sharan}, {Sullivan}, {Sumi},
  {Suzuki}, {Tristram}, {Wakiyama}, {Yonehara}, {KMTNet Modeling Team}, {Han},
  {Choi}, {Park}, {Jung}, \& {Shin}}]{strret16}
{Street}, R.~A., {Udalski}, A., {Calchi Novati}, S., {et~al.} 2016, \apj, 819,
  93

\bibitem[{{Udalski} {et~al.}(1992){Udalski}, {Szymanski}, {Kaluzny}, {Kubiak},
  \& {Mateo}}]{udalski92+04}
{Udalski}, A., {Szymanski}, M., {Kaluzny}, J., {Kubiak}, M., \& {Mateo}, M.
  1992, \actaa, 42, 253

\bibitem[{{Udalski} {et~al.}(2015){Udalski}, {Szyma{\'n}ski}, \&
  {Szyma{\'n}ski}}]{udalski15+02}
{Udalski}, A., {Szyma{\'n}ski}, M.~K., \& {Szyma{\'n}ski}, G. 2015, \actaa, 65,
  1

\bibitem[{{Wallerstein}(2002)}]{wallerstein02}
{Wallerstein}, G. 2002, \pasp, 114, 689

\bibitem[{{Wilson}(1975)}]{wilson75}
{Wilson}, C.~P. 1975, \aj, 80, 175

\bibitem[{{Wrobel} {et~al.}(2011){Wrobel}, {Greene}, \& {Ho}}]{wrobel11+02}
{Wrobel}, J.~M., {Greene}, J.~E., \& {Ho}, L.~C. 2011, \aj, 142, 113

\end{thebibliography}

\onecolumn{
\tablefirsthead{\toprule
\hline
Var & Other &   RA  &  Dec  &Epoch & $P$ & $I_{median}$ &
$A_{\mathrm{i}^{\prime}+\mathrm{z}^{\prime}}$ & $N$ & Type\\
id  & id & J2000 & J2000 & HJD  &  d  &      mag     &                mag    
                       &     &     \\  
(1) & (2) &   (3)  &  (4)  & (5) & (6) & (7) &
(8) & (9) & (10) \\
\midrule}
\tablehead{%
\multicolumn{9}{c}%
{{\bfseries  
Continued from previous page}} \\
\toprule
Var & Other &   RA  &  Dec  &Epoch & $P$ & $I_{median}$ &
$A_{\mathrm{i}^{\prime}+\mathrm{z}^{\prime}}$ & $N$ & Type\\
id & id  & J2000 & J2000 & HJD  &  d  &      mag     &                mag    
                       &     &     \\ 
(1) & (2) &   (3)  &  (4)  & (5) & (6) & (7) &
(8) & (9) & (10) \\ \midrule}
\tabletail{%
\midrule 
         \multicolumn{9}{c}{$^{a}$Epochs, periods, mean magnitudes, amplitudes, and classifications taken from OGLE database}\\
         \multicolumn{9}{c}{$^{b}$=SPVSgr18550405-3028580; $^{c}$=SPVSgr18550386-3028593; $^{d}$=Peak magnitude}\\ 
         \multicolumn{9}{c}{\textbf{Continued on next page}}\\
\midrule
}
\tablelasttail{%
\\ \midrule
         \multicolumn{9}{c}{$^{a}$Epochs, periods, mean magnitudes, amplitudes, and classifications taken from OGLE database}\\
         \multicolumn{9}{c}{$^{b}$=SPVSgr18550405-3028580; $^{c}$=SPVSgr18550386-3028593; $^{d}$=Peak magnitude}\\
         \multicolumn{9}{c}{\textbf{Concluded}} \\
\bottomrule}
\tablecaption{Ephemerides and main characteristics of the variable stars in the field of globular cluster NGC~6715. Column 1 is the id assigned to the variable
star, Column 2 is a previously known id assigned to the stars (5 digit numbers correspond to OGLE identifications of the form OGLE-BLG-RRLYR-{\it NNNNN}), Columns 3 and 4 
correspond to the right ascension and declination (J2000), Column 5 is the epoch used, Column 6 is the period measured in this work unless the variable is an OGLE star in which case we use their period, Column 7 is median magnitude, Column 8 is the peak-to-peak amplitude in the light curve, Column 9 is the number of epochs and Column 10 is the classification of the variable. The numbers in parentheses indicate the uncertainty on the last decimal place of the period.}
\label{tab:NGC6715_ephemerides}
\begin{supertabular}{cccccccccc}
V112 & -    & 18:55:02.010 & -30:28:35.13 & 2457190.7419 & 100(1) & 13.63 & 0.34 & 44 & SR \\                       
V160 & VC34; 37595 & 18:55:03.954 & -30:28:29.83 & 2457189.7725 & 0.62813716(382) & 17.65 & 0.75 & 40 & RR0 \\      
V173 & -    & 18:55:02.473 & -30:29:00.04 & 2457206.6555 & 0.360284(152) & 17.69 & 0.30 & 44 & RR1 \\
V181 & VC28 & 18:55:02.395 & -30:28:34.20 & 2457145.9107 & 0.877072(898) & 17.60 & 0.87 & 42 & RR0 \\               
V192 & VC46; 37568 & 18:55:01.973 & -30:29:02.27 & 2456784.9179 & 0.60035438(1060) & 17.63 & 0.60 & 44 & RR0 \\
\hline
V212 & VC35 & 18:55:03.777 & -30:29:03.51 & 2457214.8470 & 0.202144(48) & 17.59 & 0.79 & 42 & E \\
V213 & VC40 & 18:55:02.126 & -30:28:39.63 & 2456771.8623 & 0.286053(96)  & 17.95 & 0.64 & 44 & RR1 \\
V214 & VC39 & 18:55:02.964 & -30:28:35.37 & 2456765.9179 & 0.305386(109) & 16.97 & 0.45 & 44 & RR1 \\
V215 & VC25 & 18:55:02.701 & -30:28:52.71 & 2456789.8857 & 0.307160(110) & 17.45 & 0.34 & 44 & RR1 \\
V216 & VC9  & 18:55:04.011 & -30:28:37.73 & 2457292.6392 & 0.331073(128) & 16.49 & 0.20 & 44 & RR1 \\
V217 &  -   & 18:55:02.070 & -30:29:03.41 & 2457205.7553 & 0.331556(128) & 17.76 & 0.35 & 44 & RR1 \\
V218 &  -   & 18:55:03.556 & -30:28:56.77 & 2456891.5200 & 0.348450(142) & 16.92 & 0.20 & 44 & RR1 \\
V219 &  -   & 18:55:03.233 & -30:28:59.48 & 2457183.7770 & 0.382637(171) & 17.21 & 0.24 & 44 & RR1 \\
V220 & VC4  & 18:55:03.197 & -30:28:38.54 & 2456765.9102 & 0.388893(177) & 16.86 & 0.27 & 44 & RR1 \\
V221 & VC8  & 18:55:04.353 & -30:28:51.16 & 2456846.9179 & 0.459608(247) & 17.39 & 0.71 & 44 & RR01 \\
V222 &  -   & 18:55:02.504 & -30:28:46.40 & 2456765.9179 & 0.465653(253) & 17.70 & 0.70 & 44 & RR0? \\
V223 & VC10 & 18:55:02.425 & -30:28:45.20 & 2456823.8857 & 0.471901(260) & 17.63 & 0.81 & 44 & RR0 \\
V224 & VC20 & 18:55:03.549 & -30:28:45.09 & 2457292.6392 & 0.482336(272) & 16.16 & 0.27 & 44 & RR0? \\
V225 & VC37 & 18:55:03.862 & -30:28:43.86 & 2456464.9081 & 0.483029(272) & 17.28 & 0.35 & 44 & RR1? \\
V226 &  -   & 18:55:03.948 & -30:28:42.65 & 2456784.9179 & 0.497941(289) & 16.39 & 0.36 & 44 & RR0? \\
V227 & 37582 & 18:55:03.050 & -30:28:35.81 & 2457190.7991 & 0.50805795(495) & 17.66 & 0.90 & 44 & RR0 \\
V228 & 37575 & 18:55:02.757 & -30:28:50.88 & 2456534.6709 & 0.52522562(844) & 16.87 & 0.41 & 44 & RR0 \\
V229$^{a}$ & 37597; $^{b}$ & 18:55:04.033 & -30:28:58.20 & 2456788.9321 & 0.52679780(281) & 16.38 & 0.28 & 41 & RR0 \\
V230 &   -  & 18:55:03.637 & -30:28:55.39 & 2456465.8700 & 0.532617(331) & 16.04 & 0.29 & 44 & RR0 \\              
V231 &   -  & 18:55:02.894 & -30:28:58.30 & 2457189.8090 & 0.534194(333) & 17.80 & 0.83 & 44 & RR0 \\              
V232 & VC41 & 18:55:03.581 & -30:28:47.60 & 2456543.5260 & 0.556536(362) & 16.35 & 0.29 & 44 & RR0 \\              
V233 & VC30; 37591 & 18:55:03.735 & -30:28:40.24 & 2456846.9179 & 0.55752742(766) & 17.41 & 0.93 & 44 & RR0 \\     
V234 & VC7  & 18:55:02.255 & -30:28:37.13 & 2457206.6400 & 0.559765(366) & 17.72 & 0.60 & 44 & RR0 \\              
V235 &  -   & 18:55:03.411 & -30:29:05.75 & 2457292.6392 & 0.566711(375) & 17.41 & 0.69 & 43 & RR0 \\              
V236 & VC36; 37585 & 18:55:03.149 & -30:28:34.86 & 2457152.8500 & 0.56847125(715) & 16.53 & 0.28 & 44 & RR0 \\     
V237 & VC31; 37570 & 18:55:02.112 & -30:28:44.40 & 2456435.8882 & 0.57985290(735) & 17.57 & 0.65 & 44 & RR0 \\     
V238 &   -  & 18:55:03.242 & -30:28:42.94 & 2456846.9179 & 0.584918(399) & 17.63 & 1.69 & 44 & RR0 \\              
V239 &   -  & 18:55:02.848 & -30:29:06.91 & 2457189.8090 & 0.596179(432) & 17.57 & 0.69 & 39 & RR0 \\              
V240 & 37576 & 18:55:02.725 & -30:28:30.46 & 2456909.5275 & 0.59629933(890) & 17.06 & 0.54 & 43 & RR0 \\           
V241 & VC1  & 18:55:04.190 & -30:28:39.15 & 2456896.5087 & 0.602588(424) & 17.79 & 0.65 & 44 & RR0 \\              
V242 &   -  & 18:55:02.756 & -30:29:00.15 & 2456891.4700 & 0.604762(427) & 16.23 & 0.16 & 44 & RR0 \\              
V243 &   -  & 18:55:02.587 & -30:28:38.62 & 2456770.8691 & 0.610170(435) & 16.97 & 0.36 & 44 & RR0 \\              
V244$^{a}$ & 37581 & 18:55:03.031 & -30:29:01.15 & 2456891.4945 & 0.61517949(819) & 16.90 & 0.27 & 44 & RR0 \\     
V245 & VC26 & 18:55:02.849 & -30:28:57.81 & 2456771.8623 & 0.626797(459) & 17.68 & 0.47 & 44 & RR0 \\              
V246$^{a}$ & 37573 & 18:55:02.616 & -30:28:40.91 & 2457207.8691 & 0.62879467(894) & 16.11 & 0.06 & 42 & RR0 \\     
V247 &   -  & 18:55:02.826 & -30:28:43.51 & 2456846.9179 & 0.650946(495) & 16.86 & 0.42 & 44 & RR0 \\              
V248 &   -  & 18:55:03.795 & -30:28:54.42 & 2457258.7147 & 0.668967(522) & 17.05 & 0.46 & 44 & RR0 \\              
V249 &   -  & 18:55:03.226 & -30:28:43.99 & 2457191.7238 & 0.673703(530) & 15.91 & 0.21 & 44 & RR0 \\              
V250 & VC23; 37590 & 18:55:03.705 & -30:29:00.36 & 2456543.5260 & 0.68064951(790) & 17.54 & 0.58 & 44 & RR0 \\     
V251 & 37579 & 18:55:02.876 & -30:28:30.48 & 2457189.7725 & 0.68928616(716) & 17.54 & 0.88 & 43 & RR0 \\           
V252 & 37593 & 18:55:03.792 & -30:29:07.83 & 2457189.8090 & 0.72866844(702) & 17.28 & 0.50 & 17 & RR0 \\           
V253 & 37586 & 18:55:03.268 & -30:28:28.69 & 2457206.6555 & 0.74391157(1160) & 17.32 & 0.29 & 17 & RR0 \\          
V254 &   -   & 18:55:02.808 & -30:28:49.69 & 2456436.8969 & 0.747628(652) & 17.06 & 0.71 & 44 & RR0 \\             
V255 & VC5  & 18:55:03.123 & -30:28:37.99 & 2456765.8300 & 0.760088(674) & 16.48 & 0.19 & 44 & RR0 \\              
V256 &   -  & 18:55:03.730 & -30:28:44.94 & 2456784.9179 & 14.771(25) & 14.72 & 0.71 & 44 & CWA \\                
V257 &   -  & 18:55:02.996 & -30:28:43.18 & 2456823.8857 & 20.747(50) & 14.04 & 0.09 & 44 & SR \\                 
V258 &   -  & 18:55:03.732 & -30:28:48.34 & 2457224.5000 & 37.980(168) & 13.16 & 0.04 & 44 & SR \\                 
V259 &   -  & 18:55:04.216 & -30:28:49.91 & 2456464.9081 & 154(3) & 13.30 & 0.45 & 44 & SR \\                      
V260 &   -  & 18:55:03.170 & -30:28:54.44 & -- & -- & 12.76 & 0.06 & 44 & L \\                                     
V261 &   -  & 18:55:03.399 & -30:28:44.99 & -- & -- & 13.78 & 0.07 & 44 & L \\                                     
V262 &   -  & 18:55:01.876 & -30:28:43.79 & -- & -- & 13.99 & 0.07 & 40 & L \\                                     
V263 &  $^{c}$  & 18:55:03.877 & -30:28:59.31 & -- & -- & 14.71 & 0.07 & 44 & L \\                                     
V264 &   -  & 18:55:03.753 & -30:28:49.36 & -- & -- & 13.87 & 0.08 & 44 & L \\                                     
V265 &   -  & 18:55:02.987 & -30:28:51.96 & -- & -- & 14.00 & 0.08 & 44 & L \\
V266 &   -  & 18:55:03.458 & -30:28:46.47 & -- & -- & 14.04 & 0.08 & 44 & L \\
V267 &   -  & 18:55:03.463 & -30:28:45.05 & -- & -- & 13.66 & 0.09 & 44 & L \\
V268 &   -  & 18:55:02.998 & -30:28:44.30 & -- & -- & 13.79 & 0.09 & 44 & L \\
V269 &   -  & 18:55:03.712 & -30:28:43.08 & -- & -- & 13.62 & 0.10 & 44 & L \\
V270 &   -  & 18:55:02.603 & -30:28:48.92 & -- & -- & 13.70 & 0.11 & 44 & L \\
V271 &   -  & 18:55:04.004 & -30:28:37.05 & -- & -- & 13.97 & 0.12 & 44 & L \\
V272 &   -  & 18:55:03.065 & -30:28:47.99 & -- & -- & 14.25 & 0.05 & 44 & L \\
V273 &   -  & 18:55:03.708 & -30:28:35.30 & -- & -- & 13.40 & 0.15 & 44 & L \\
V274 &   -  & 18:55:03.833 & -30:29:03.89 & -- & -- & 13.47 & 0.08 & 44 & L \\
V275 &   -  & 18:55:02.346 & -30:28:43.40 & -- & -- & 13.67 & 0.16 & 44 & L \\
V276 &   -  & 18:55:03.659 & -30:28:50.52 & -- & -- & 14.06 & 0.16 & 44 & L \\
V277 &   -  & 18:55:03.309 & -30:28:49.79 & -- & -- & 13.71 & 0.19 & 44 & L \\
V278 &   -  & 18:55:04.070 & -30:28:49.79 & -- & -- & 13.13 & 0.33 & 44 & L \\
V279 &   -  & 18:55:04.045 & -30:28:41.41 & -- & -- & 13.42 & 0.34 & 44 & L \\
V280 &   -  & 18:55:03.692 & -30:28:52.01 & -- & -- & 13.25 & 0.46 & 44 & L \\
V281 &   -  & 18:55:03.324 & -30:28:47.26 & -- & -- & 16.71 & 0.45 & 44 & NC \\
V282 &   -  & 18:55:03.010 & -30:28:35.82 & -- & -- & 17.52 & 0.74 & 44 & NC \\
V283 &   -  & 18:55:03.747 & -30:28:53.88 & -- & -- & 16.97 & 0.76 & 44 & NC \\
V284 & VC32 & 18:55:03.367 & -30:28:37.97 & -- & -- & 17.14 & 0.94 & 44 & NC \\
V285 & VC33 & 18:55:02.686 & -30:28:32.48 & -- & -- & 17.52 & 0.55 & 44 & NC \\
V286 &   -  & 18:55:03.536 & -30:28:51.06 & -- & -- & 15.04$^{d}$ & -- & 43 & NC \\
V287 &   -  & 18:55:03.373 & -30:28:35.86 & -- & -- & -- & -- & 44 & NC \\
V288 &   -  & 18:55:03.412 & -30:28:52.76 & -- & -- & -- & -- & 44 & NC \\
V289 &   -  & 18:55:02.904 & -30:28:39.89 & -- & -- & -- & -- & 42 & NC \\
V290 &   -  & 18:55:03.475 & -30:29:06.23 & -- & -- & -- & -- & 39 & NC \\
V291 & VC27 & 18:55:03.258 & -30:28:41.09 & -- & -- & -- & -- & 42 & NC \\

\end{supertabular}%
}
\twocolumn

\begin{figure*}
\centering
\subfloat[][]{\includegraphics[scale=1.0]{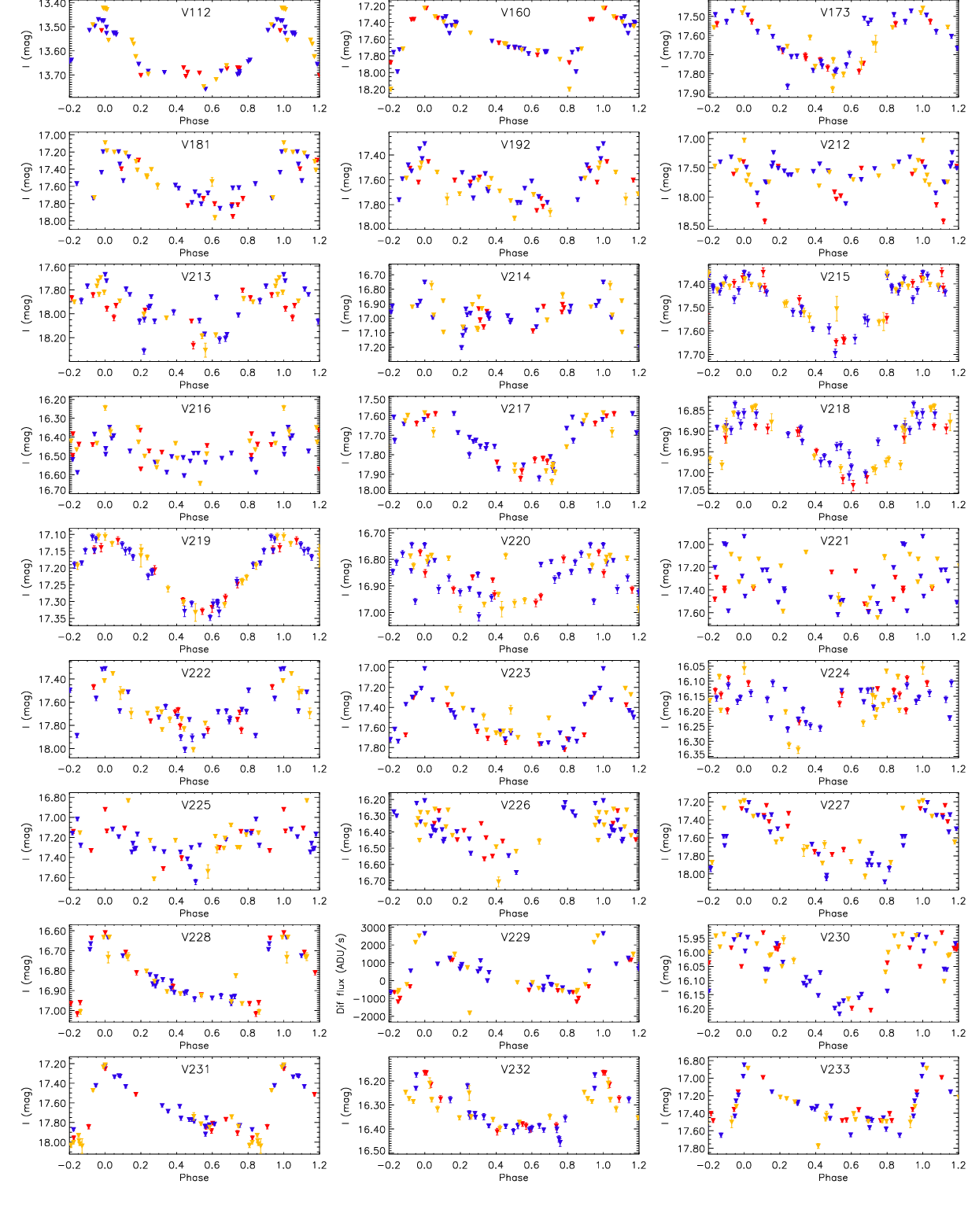}}
\caption{Light curves of the known and new variables discovered in globular cluster NGC~6715. Red, blue, and yellow triangles correspond to the data obtained during the years 2013, 2014, and 2015, respectively. For V229, V244, V246, V286-V291, we plot the quantity $f_{\mathrm{diff}}(t)/p(t)$ since a reference flux is not available.}
\label{fig:lc_variables}
\end{figure*}

\begin{figure*}
\ContinuedFloat
\centering
\subfloat[][]{\includegraphics[scale=1.0]{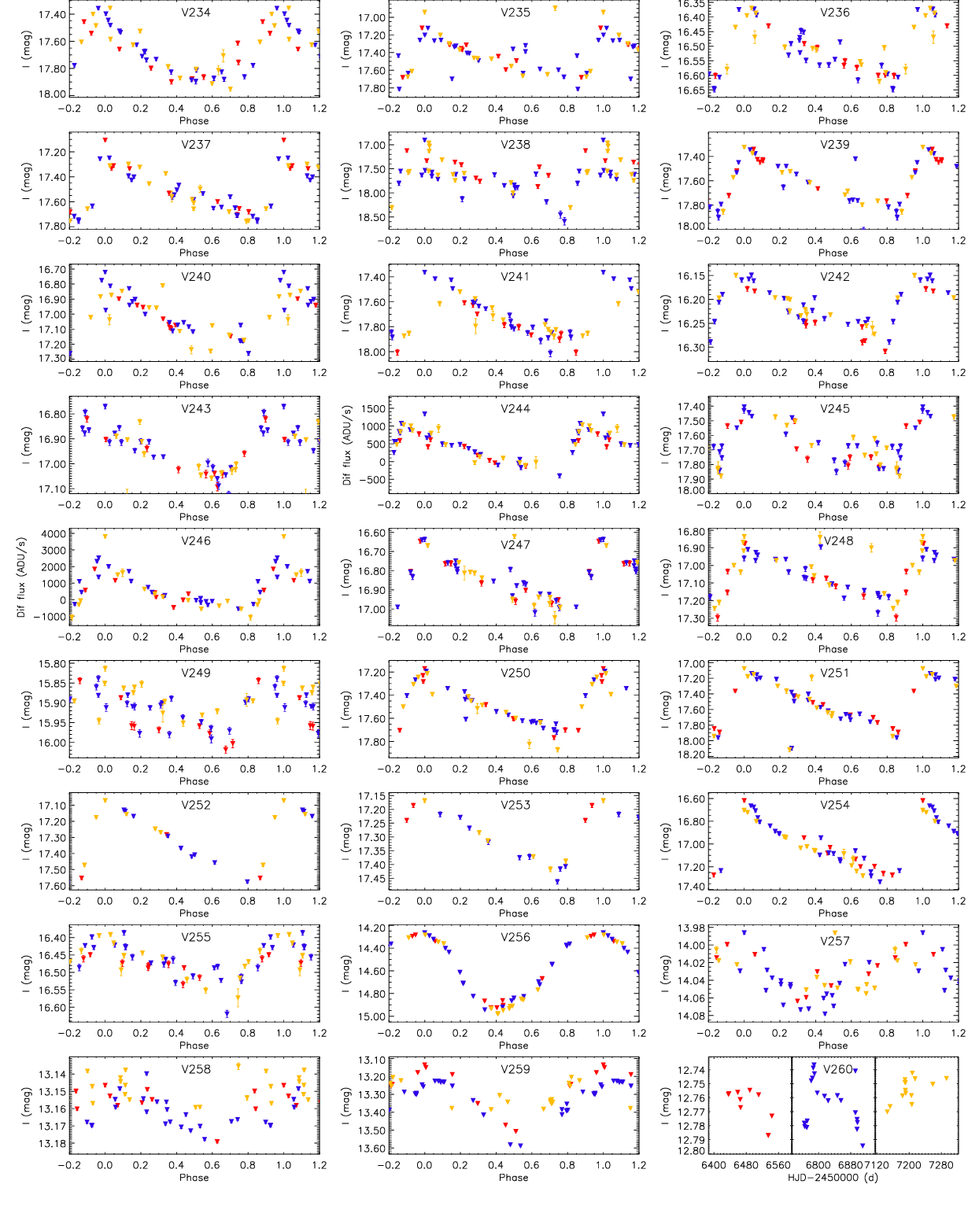}}
\caption{Light curves of the known and new variables discovered in globular cluster NGC~6715. Red, blue, and yellow triangles correspond to the data obtained during the years 2013, 2014, and 2015, respectively. For V229, V244, V246, V286-V291, we plot the quantity $f_{\mathrm{diff}}(t)/p(t)$ since a reference flux is not available \emph{(cont.)}.}
\end{figure*}

\begin{figure*}
\ContinuedFloat
\centering
\subfloat[][]{\includegraphics[scale=1.0]{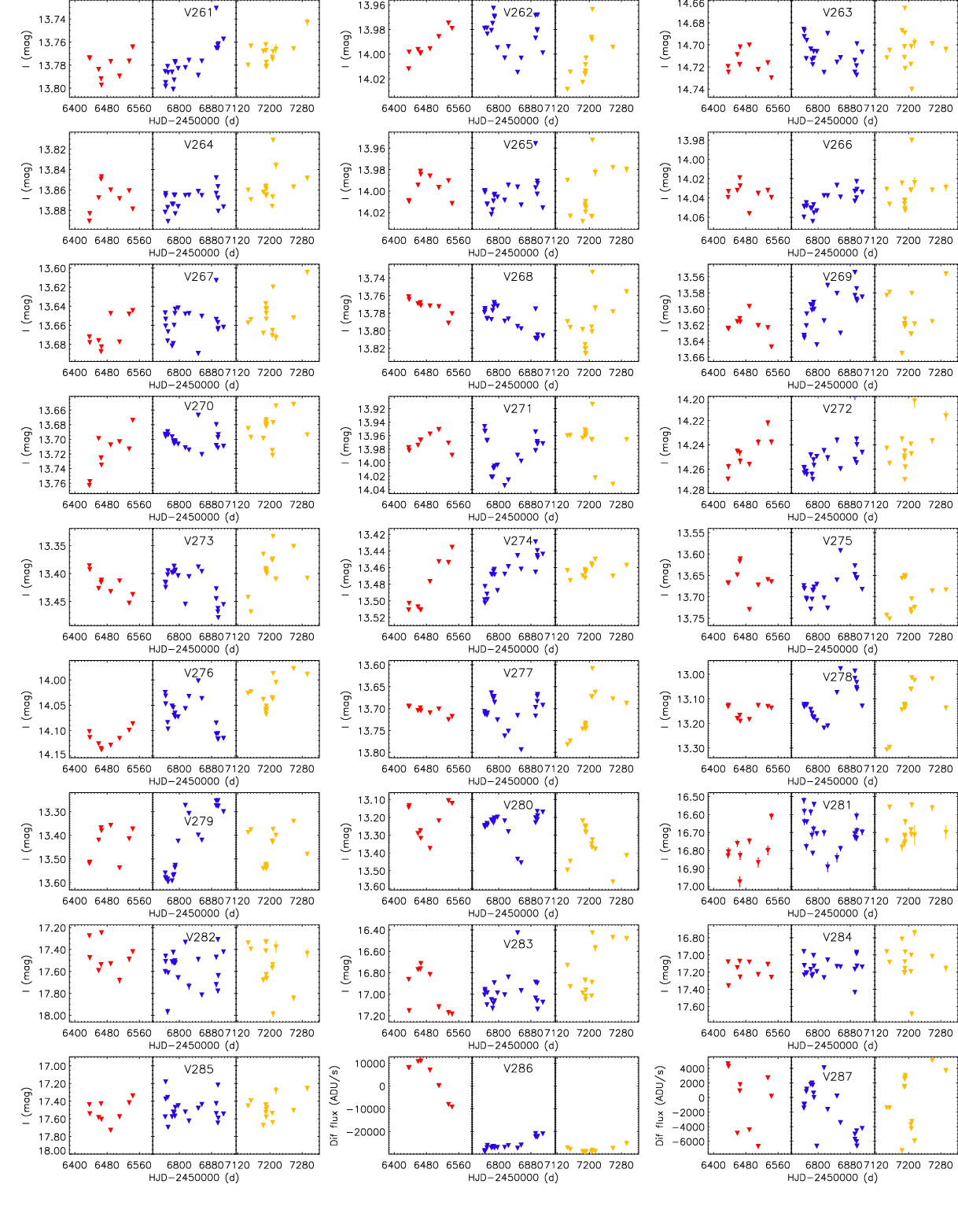}}
\caption{Light curves of the known and new variables discovered in globular cluster NGC~6715. Red, blue, and yellow triangles correspond to the data obtained during the years 2013, 2014, and 2015, respectively. For V229, V244, V246, V286-V291, we plot the quantity $f_{\mathrm{diff}}(t)/p(t)$ since a reference flux is not available \emph{(cont.)}.}
\end{figure*}

\begin{figure*}
\ContinuedFloat
\centering
\subfloat[][]{\includegraphics[scale=1.0]{eps_plots/lc_chart4_phase_hjd.eps}}
\caption{Light curves of the known and new variables discovered in globular cluster NGC~6715. Red, blue, and yellow triangles correspond to the data obtained during the years 2013, 2014, and 2015, respectively. For V229, V244, V246, V286-V291, we plot the quantity $f_{\mathrm{diff}}(t)/p(t)$ since a reference flux is not available \emph{(cont.)}.}
\end{figure*}

\begin{appendix}

\section{Cross identification between variables}\label{apen:cross_id}
\begin{table*}[htp!]
\caption{Cross identification between the RR Lyrae variables listed by the CVSGGC \citep{clement01}, the variable star candidates from \cite{montiel10+01}, and the OGLE RR Lyrae stars \citep{udalski15+02}}
\label{tab:ogle_montiel_cross_id}
\centering
\begin{tabular}{ccccccccc}
\hline\hline
CVSGGC & OGLE & Montiel & CVSGGC & OGLE & Montiel & CVSGGC & OGLE & Montiel \\
id     &  id     & id   & id     &  id     & id   & id     &  id     & id   \\
       & (OGLE-BLG-) &        &         &      &        &         &     &    \\
\hline
V3   & RRLYR-37571 &      & V92  & RRLYR-37611 &      & V186 & RRLYR-37546 &      \\
V4   & RRLYR-37540 &      & V93  & RRLYR-37567 &      & V188 & RRLYR-37584 &      \\
V5   & RRLYR-37518 &      & V94  & RRLYR-37634 &      & V192 & RRLYR-37568 & VC46 \\
V6   & RRLYR-37655 &      & V95  & RRLYR-37560 & VC13 & V193 & RRLYR-37605 &      \\
V7   & RRLYR-37620 &      & V97  & RRLYR-37607 &      & V194 & RRLYR-37574 &      \\
V12  & RRLYR-37511 &      & V98  & RRLYR-37637 &      & V227 & RRLYR-37582 &      \\
V15  & RRLYR-37649 &      & V118 & RRLYR-37660 &      & V228 & RRLYR-37575 &      \\
V28  & RRLYR-37623 &      & V119 & RRLYR-37654 &      & V229 & RRLYR-37597 &      \\
V29  & RRLYR-37519 &      & V120 & RRLYR-37638 &      & V233 & RRLYR-37591 &      \\
V31  & RRLYR-37528 &      & V121 & RRLYR-37635 &      & V236 & RRLYR-37585 &      \\
V32  & RRLYR-37514 &      & V122 & RRLYR-37632 &      & V237 & RRLYR-37570 &      \\
V33  & RRLYR-37631 &      & V123 & RRLYR-37583 &      & V240 & RRLYR-37576 &      \\
V34  & RRLYR-37541 &      & V124 & RRLYR-37628 &      & V244 & RRLYR-37581 &      \\
V35  & RRLYR-37533 &      & V125 & RRLYR-37613 &      & V246 & RRLYR-37573 &      \\
V36  & RRLYR-37648 &      & V126 & RRLYR-37600 &      & V250 & RRLYR-37590 &      \\
V37  & RRLYR-37617 &      & V127 & RRLYR-37580 & VC2  & V251 & RRLYR-37579 &      \\
V38  & RRLYR-37538 &      & V128 & RRLYR-37551 &      & V252 & RRLYR-37593 &      \\
V39  & RRLYR-37527 &      & V129 & RRLYR-37526 & VC17 & V253 & RRLYR-37586 &      \\
V40  & RRLYR-37543 &      & V130 & RRLYR-37549 &      & --   & RRLYR-37565 &      \\
V41  & RRLYR-37647 &      & V131 & RRLYR-37525 &      & --   & RRLYR-37578 & VC38 \\
V42  & RRLYR-37626 &      & V132 & RRLYR-37555 &      & --   & RRLYR-37594 &      \\
V43  & RRLYR-37516 &      & V133 & RRLYR-37550 &      & --   & RRLYR-37621 &      \\
V44  & RRLYR-37599 &      & V136 & RRLYR-37513 &      &      &             &      \\
V45  & RRLYR-37646 &      & V137 & RRLYR-37612 &      &      &             &      \\
V46  & RRLYR-37552 & VC44 & V138 & RRLYR-37610 &      &      &             &      \\
V47  & RRLYR-37554 &      & V139 & RRLYR-37596 &      &      &             &      \\
V48  & RRLYR-37659 &      & V140 & RRLYR-37572 &      &      &             &      \\
V49  & RRLYR-37532 &      & V141 & RRLYR-37558 &      &      &             &      \\
V50  & RRLYR-37640 &      & V142 & RRLYR-37545 & VC15 &      &             &      \\
V51  & RRLYR-37657 &      & V143 & RRLYR-37534 &      &      &             &      \\
V52  & RRLYR-37636 &      & V147 & RRLYR-37602 &      &      &             &      \\
V54  & RRLYR-37521 &      & V148 &     --      & VC45 &      &             &      \\
V55  & RRLYR-37651 &      & V151 & RRLYR-37517 &      &      &             &      \\
V57  & RRLYR-37666 &      & V152 & RRLYR-37615 &      &      &             &      \\
V58  & RRLYR-37630 &      & V158 & RRLYR-37539 &      &      &             &      \\
V59  & RRLYR-37512 &      & V159 & RRLYR-37589 &      &      &             &      \\
V60  & RRLYR-37509 &      & V160 & RRLYR-37595 & VC34 &      &             &      \\
V61  & RRLYR-37547 &      & V161 & RRLYR-37616 &      &      &             &      \\
V62  & RRLYR-37531 &      & V162 & RRLYR-37592 & VC11 &      &             &      \\
V63  & RRLYR-37553 &      & V163 & RRLYR-37604 & VC12 &      &             &      \\
V67  & RRLYR-37587 &      & V164 & RRLYR-37564 & VC14 &      &             &      \\
V69  & RRLYR-37505 &      & V165 & RRLYR-37609 &      &      &             &      \\
V74  & RRLYR-37645 &      & V168 & RRLYR-37619 &      &      &             &      \\
V76  & RRLYR-37529 & VC47 & V172 & RRLYR-37614 &      &      &             &      \\
V77  & RRLYR-37524 &      & V174 & RRLYR-37608 &      &      &             &      \\
V78  & RRLYR-37629 &      & V176 & RRLYR-37618 &      &      &             &      \\
V80  & RRLYR-37624 &      & V177 & RRLYR-37556 &      &      &             &      \\
V82  & RRLYR-37548 &      & V178 & RRLYR-37562 &      &      &             &      \\
V83  & RRLYR-37522 &      & V179 & RRLYR-37557 & VC18 &      &             &      \\
V84  & RRLYR-37577 &      & V180 & RRLYR-37588 &      &      &             &      \\
V85  & RRLYR-37569 &      & V181 &     --      & VC28 &      &             &      \\
V87  & RRLYR-37641 &      & V182 & RRLYR-37561 &      &      &             &      \\
V88  & RRLYR-37515 &      & V183 & RRLYR-37603 &      &      &             &      \\
V89  & RRLYR-37530 &      & V184 & RRLYR-37559 &      &      &             &      \\
V90  & RRLYR-37622 &      & V185 & RRLYR-37563 &      &      &             &      \\

\hline
\end{tabular}
\end{table*}

\end{appendix}

\end{document}